\documentclass{aa}
\usepackage{amsmath,bm} %amsthm
\usepackage[varg]{txfonts}
\usepackage{natbib}
\usepackage{chemformula}

\usepackage{afterpage}
\usepackage[markup=bfit]{changes}
\usepackage[colorlinks=true,linkcolor=blue,citecolor=blue, urlcolor=blue]{hyperref}
\bibpunct{(}{)}{;}{a}{}{,}

\usepackage{pdflscape}
\usepackage{afterpage}
\usepackage{mathtools}

\usepackage{booktabs}
\usepackage{gensymb}

\usepackage{color}
\usepackage{natbib,twoopt}
\usepackage[hyphenbreaks]{breakurl}

\begin{document} 
   \title{The ionization structure and chemical history in isolated \ion{H}{ii} regions of dwarf galaxies with IFU}
   \subtitle{I. The Sagittarius Dwarf Irregular Galaxy\thanks{Based on observations from the ESO program ID 63.N-0726(B), and 077.C-0812(A)}}
   
   \author{A. Andrade \inst{1,2} 
            \and I. Saviane\inst{2} 
            \and L. Monaco \inst{3,4}
            \and M. Gullieuszik \inst{5}
          }

   \institute{Universidad Andres Bello, Facultad de Ciencias Exactas, Departamento de Física y Astronomía - Instituto de Astrofísica, Fernández Concha 700, Las Condes, Santiago, Chile. \\   
   \email{a.andradevalenzuela$@$uandresbello.edu}
    \and
        European Southern Observatory, Alonso de Cordova 3107, Vitacura, Casilla 19001, Santiago de Chile 19, Chile.
    \and 
        Universidad Andres Bello, Facultad de Ciencias Exactas, Departamento de Física y Astronomía - Instituto de Astrofísica, Autopista
        Concepción-Talcahuano, 7100, Talcahuano, Chile.
    \and 
        INAF – Osservatorio Astronomico di Trieste, Via G. B. Tiepolo 11, 34143 Trieste, Italy
    \and 
        INAF - Osservatorio Astronomico di Padova, Vicolo dell’Osservatorio 5, I-35122 Padova, Italy.}

   \date{Received January 03, 2025; Accepted  May 22, 2025 (A$\&$A)}
 % \abstract{}{}{}{}{} 
% 5 {} token are mandatory
  \abstract
  {Studying metal-poor galaxies is crucial for the understanding of physical mechanisms that drive the formation and evolution of galaxies, such as internal dynamics, star formation history, and chemical enrichment. Most of the observational works in dwarf galaxies employ integral field spectroscopy to investigate gas physics in the entire body of galaxies. \textcolor{black}{However, these past studies have not investigated the detailed spatially resolved properties of individual extragalactic \ion{H}{ii} regions.}}
  {We study the only known \ion{H}{ii} region in the Sagittarius dwarf irregular galaxy, a metal-poor galaxy of the local universe, using IFU VIMOS/VLT and long-slit EFOSC2/NTT archival data. We explore the spatially resolved gaseous structure by using optical emission lines, to (i) give insights into the physical processes that are shaping the evolution of this \ion{H}{ii} region, and (ii) relate these mechanisms to the metal-poor gas-phase component in extragalactic \ion{H}{ii} regions.}
  {We probe optical emission line structures of the \ion{H}{ii} region, fully covered within the $27''\times 27''$ field of view of VIMOS. the oxygen abundances were estimated by applying the $T_{e}-$sensitive method, by using the auroral $[\ion{O}{iii}]\lambda4363$ emission line detectable at S/N>3 integrating the spectral fibers of the data cube.}
  {From emission line maps, the $\mathrm{O^{++}}$ emission is concentrated towards the centre, in comparison with the low ionized species such as $\mathrm{O^{+}}$, and $\mathrm{H^{+}}$. The H$\beta$ maps reveal that the \ion{H}{ii} shows two prominent clumps, showing a biconic-like shape aligned along the same axis. Radial flux-density profiles reveal that those clumps are similar in terms of size ($\sim$8$''$) and flux distribution in H$\beta$ and $\mathrm{[\ion{O}{iii}]\lambda5007}$. Comparing stellar populations from HST photometry in the gaseous structure, we found that old stellar populations ($>1$ Gyr) are uniformly distributed across the \ion{H}{ii} region, whereas the young stellar populations ($\lessapprox 700$ Myr) are found closer to the edges of the H$\beta$ clumps, distributed in filamentary configurations. \textcolor{black}{We estimate $T_{e}= 17683 \pm 1254$ K for the gaseous structure.} The $T_{e}-$based oxygen abundance of the SagDIG \ion{H}{ii} region is \textcolor{black}{$\mathrm{12+log(O/H) = 7.23\pm0.04}$}, which is in agreement with empirical estimations of the literature, and is also in line with the low-mass end of the mass-metallicity relation  (MZR). \textcolor{black}{Considering corrections on $T_{e}$ fluctuations, we estimate $\mathrm{12+log(O/H) = 7.50\pm0.08}$}.}
  {The stratified composition of the \ion{H}{ii} region is a signature that this gaseous structure is expanding. This feature, together with SagDIG falling in the low-mass end of the mass-metallicity relation, suggests that the evolution of this \ion{H}{ii} region is being sustained by ionisation from massive stars, stellar winds, and supernovae explosions, expanding the gas structure. The filamentary configuration of young stars is likely produced by the interaction between atomic and ionized gas, being in line with many galactic \ion{H}{ii} regions and those found in the Large Magellanic Cloud. If this proposed scenario is confirmed with multi-wavelength data and data cubes with better spectral coverage and spatial resolution, it could implicate that \ion{H}{ii} regions in metal-poor dwarf galaxies are subject to the same physics as \ion{H}{ii} regions in the Milky-Way.}
    %{}


   \maketitle
% ---------------------------------------------------------------------------
\section{Introduction}

The local universe has been our preferred laboratory for understanding the formation and evolution of dwarf galaxies. These systems are dominating in number the universe (\citealt{schechter76}, \citealt{moffett16}), and present a huge variety in terms of their structure and kinematics, such as dwarf irregular (dIrr) galaxies, dwarf spheroidal (dSph) galaxies, and blue compact dwarfs (BCD). In particular, dIrrs show irregular morphologies in their gas and stellar components and are characterized by their small sizes (few hundred parsecs) and low masses (\citealt{hunter24}). Because of the shallower potential wells of dIrrs, the atomic species created by stellar nucleosynthesis are not only redistributed in the inster-stellar medium (ISM), but large amounts of metals are expelled into their haloes.
\\
In general, dIrrs have a high gas reservoir compared with their stellar masses. \ion{H}{i} observations and simulations show that the atomic gas is found in their outer regions, surrounding the stellar body (\citealt{younglo96},\citeyear{younglo97}, \citealt{hunter19}), which is decomposed in two dispersion-dominated gas velocity \ion{H}{i} profiles: a narrow ($\sim$2$-5$ km s$^{-1}$), and a broad ($\sim$10 km s$^{-1}$) component, because stellar feedback from massive stars creates thermal and material motion as turbulence, which suppresses the gas accretion with high angular momentum (\citealt{dutta09}, \citealt{bukhart10}, \citealt{oh15}, \citealt{mcnichols16}, \citealt{maier17}, \citealt{el-badry18}).
\\
Theoretical models and cosmological simulations under the $\mathrm{\Lambda CDM}$ framework have demonstrated that these systems are key to the hierarchical growth of star-forming galaxies (\citealt{tosi03}, \citealt{tolstoy09}). For this reason, their properties are usually linked to the primordial conditions of the universe (\citealt{izotov04}), and it is believed that they are the primary agent of the reionisation in the early universe (\citealt{bunker10}, \citealt{atek24})\\
\\

Dwarf galaxies have been extensively studied in terms of their chemical evolution, setting a shallower slope in the low mass end of the mass-metallicity relation (MZR), with respect to typical star-forming (SF) galaxies ($\geq 10^{9}M_{\odot}$, \citealt{lee06}, \citealt{saviane08}, \citealt{berg12}, \citealt{andrewsmartini13}, \citealt{torrey19}, \citealt{curti20}).\\

Due to shallower potential wells of dwarf galaxies, supernovae (SNe) explosions and galactic winds can accelerate enough gas particles, surpassing the escape velocity and expelling their gas-phase component, so dwarf galaxies are mostly dominated by outflows as a product of the stellar feedback, which can modify even the morphology of these systems (\citealt{tremonti04}, \citealt{tissera05}, \citealt{scannapieco08}). However, metal-poor gas accretion from galaxy-galaxy interactions or the circum-galactic medium can also dilute the gas-phase metallicity, disturbing also the galactic morphology (\citealt{montuori10}, \citealt{finlatordave08}, \citealt{perezdiaz24}). Therefore, attributing to a single physical process the chemical evolution of a galaxy is difficult, because such objects are not always in equilibrium between star formation and material flows (inflows and outflows) mechanisms (\citealt{dave11}, \citealt{lin23}).\\

Most of these works have studied physical processes inferred from chemical abundances obtained by using either long-slit spectroscopy or multi-object spectroscopy (together with masses obtained from stellar photometry or spectral fitting), because (i) the angular dimensions are not large enough to resolve the gas-phase component of distant metal-poor galaxies, and (ii) observational works in spatially-resolved metal-poor galaxies are mostly focused in analyzing the stellar populations that they harbour.
For this reason, understanding the behaviour in terms of the gas-phase metal content of different physical mechanisms, just by employing an optical spectral analysis is challenging. So employing a multi-wavelength spectro-photometric exploration of dwarf irregular galaxies can reveal important clues to the role of feedback mechanisms in their \ion{H}{ii} regions. Therefore, we put our attention to one of the most studied dwarf irregular galaxy in the local universe, the Sagittarius Dwarf Irregular Galaxy (SagDIG, $M_{*}=1.8\pm0.5\times 10^{6}M_{\odot}$, $r_{eff}= 227º\pm20$ pc, $D=1068\pm88$ Kpc; \citealt{mcconnachie12}, \citealt{kirby18}), which offers a spatially resolved stellar population, as well as \ion{H}{i} and \ion{H}{ii} spatially resolved components, compared to other dwarf irregular galaxies in the local universe.
\\

SagDIG was discovered using photometric plates obtained with the ESO Schmidt telescope by \cite{cesarsky77}. \cite{younglo97} using VLA observations probed the \ion{H}{i} medium and found that the stellar body of SagDIG is surrounded by the atomic gas.

Up to date, SagDIG has been one of the most studied extragalactic sources in terms of its stellar populations. \citeauthor{momany02} (\citeyear{momany02}, \citeyear{momany05}) using HST photometry found a well-defined red-giant branch (RGB), showing that the galaxy is dominated by populations older than 1 Gyrs, with a stellar metallicity of $\mathrm{[Fe/H]\simeq -2.0}$ dex. Moreover, the detection of carbon-rich and oxygen-rich stars in the asymptotic giant branch, and horizontal branch stars (\citealt{cook87}, \citealt{demers02}, \citealt{gullieuszik07}, \citealt{momany14}) suggests that SagDIG experienced its first star formation episode $9-10$ Gyrs ago, which lasted $\sim$8 Gyrs. On the other hand, it is also detected a well-defined blue main sequence, containing young stars with ages of a few hundred Myrs, located mainly closer to the edges of the high density \ion{H}{i} columns and remaining regions of propagated star formation.\\
\\
The life path of dIrrs has been studied from theoretical models and cosmological simulations, supported by observational evidence. In general, the common star formation history (SFH) of dIrrs shows extended episodes of star formation which are suppressed at early epochs, and in many cases, followed by a reignition of star formation to the present-day (\citealt{benitez15}, \citealt{ledinauskas18}). This is also the case of SagDIG. \cite{held07} derived the SagDIG's SFH being in line with the analysis of stellar content of \citep{momany05}: SagDIG experienced an extended period of star formation between $\sim$2-9 Gyrs ago, where the highest efficiency was reached $\sim$6 Gyr ago. Moreover, at the present day, SagDIG is showing an ongoing star-formation episode that started $\sim$1 Gyr ago, being at least 2 times stronger than the extended one. This picture is consistent with a recent SFH derived from Long-period Variable stars (LPV) by \cite{parto23}, suggesting that between $30\% - 50\%$ of the stellar mass of SagDIG was assembled in the peak of the extended star-formation episode. \\
\\
In line with the current picture of the evolution of dwarf irregular galaxies, SagDIG seems to have experienced a poor chemical enrichment. Optical spectroscopic works done on the only known \ion{H}{ii} region, by \citeauthor{skillman89a} (\citeyear{skillman89a}a,\citeyear{skillman89b}b) and \citeauthor{saviane02} (\citeyear{saviane02}, hereafter S02) using "strong-line" methods, derived a total oxygen abundance of  $7.26<\mathrm{12+log(O/H)}<7.50$, showing its low gas-phase metallicity content. However, it is well known that strong-line methods are biased to the set of the emission lines used to construct the correlation of the emission line ratios and the metallicity, showing $\sim$1 dex of difference in the most extreme cases \textcolor{black}{between different strong-line methods} (\citealt{kewley08}, \citealt{poetrodjojo21}).\\
\\
\textcolor{black}{Numerous works have presented exciting results on star-forming galaxies through IFU data (e.g. \citeauthor{james10} \citeyear{james10}, \citeyear{james13}, \citeyear{james20}, \citealt{perez-montero11}, \citealt{vanzi11}, \citeauthor{kumari17} \citeyear{kumari18}, \citealt{emsellem22}), aimed to understand the gas physics on the entire body of disk galaxies, BCGs, dIrrs, and merger systems. However, there is no reported literature to study spatially resolved gas components of individual \ion{H}{ii} regions for extragalactic sources beyond the Magellanic clouds.} For this reason, in this work, we present a new perspective on the chemical analysis in metal-poor dwarf galaxies in the local universe, by studying the spatially-resolved extragalactic \ion{H}{ii} region of SagDIG with IFU data. We aim to probe the physical structure of the SagDIG \ion{H}{ii} region and estimate the total oxygen abundance with the so-called $T_{e}-$sensitive method, the best metallicity estimator.

This paper is structured as follows: In section 2, we present archival long-slit and IFU observations together with their corresponding data reduction.  In section 3, we describe the data analysis, which includes emission line maps, flux density profiles, metallicity estimations, and a discussion about the biases of the method. In section 4, we compare the stellar population in the SagDIG \ion{H}{ii} region with its gas-phase content, and we place SagDIG in the mass-metallicity plane, discussing our results. Finally, in section 5 we present our conclusions.

\section{Observations and data reduction}
\subsection{VIMOS data}
\label{vimos_data}
We used archival data of the only known \ion{H}{ii} region in SagDIG. This gaseous nebula is located at $\alpha = 19^{h}30^{m}02.59^{s}, \ \delta=-17^{o}41'28.6''$; J2000. The data was obtained with the VIsible MultiObject Spectrograph (VIMOS, \citealt{lefevre}) under program 077.C-0812(A) in April 2006 (PI: M. Gullieuszik). VIMOS was a visible (360 nm to 1000 nm) wide-field imager and multi-object spectrograph mounted on Nasmyth focus B of VLT/UT 3 (Melipal). The Integral Field Unit (IFU) mode was used to explore the spatially resolved gaseous nebula in three observing blocks (OBs) of one hour each.\\

The IFU consists of 1600 fibres (pseudo-slits) storing 400 fibres per quadrant. The field of view (FoV) covers a sky region of $27''\times 27''$ with a spatial resolution of $0.67''\ \mathrm{px}^{-1}$ in the wide-field mode. This is enough to observe completely the gaseous nebula and also an adjacent galaxy field free from nebular emission, so a significant number of fibres were used to represent the sky spectra for the respective decontamination of telluric lines and sky subtraction. Observations were conducted with the HR-blue grism which covers the wavelength domain of $4150-6200 \AA$, with a spectral resolution of  $R = 2020\ (0.51\AA\ \mathrm{px}^{-1})$. We detected the Balmer emission lines H$\beta$ and H$\gamma$, the collisional excitation $[\ion{O}{iii}]\lambda\lambda 4959,5007$ emission line doublet, all of them with $\mathrm{S/N>3}$, allowing us to study the physical structure of the gaseous nebula.\\

Each frame was reduced independently with the VIMOS Pipeline in the EsoReflex environment (\citealt{esoreflex}). The pipeline performs bias subtraction, flat normalisation, wavelength calibration, and flux calibration. For the latter, the calibration was done by using spectrophotometric standard stars from \citeauthor{hamuy92} (\citeyear{hamuy92}, \citeyear{hamuy94}) and \citet{moehler14}: The F-type LTT3864 ($V = 12.17, \ B-V =0.50 $), the G-type LTT7379 ($V = 10.23, \ B-V = 0.61$), and the DA-type EG247 ($V = 11.03,\  B-V = -0.14$). The final products are flux-calibrated in units of $10^{-16}\mathrm{erg\ s^{-1}\ cm^{-2}\ \AA^{-1}}$. Cosmic rays were removed by taking the median of the three spectra per spaxel (one per OB). Those spaxels that have two available spectra (because of the exposure superposition and bad pixels) were averaged to represent the main spectrum per spaxel. 

In order to reduce the spectral noise in each spectral fibre, we apply an average fibre smoothing. This technique works by weighting each spectral bin with the mean values of its neighbor bins. Three neighbor bins per spectral bin were used to apply the smoothing. A second-order Chebyshev polynomial was fitted by using Specultils (\citealt{specutils}) to subtract the continuum contribution.

\subsection{The auroral $\mathrm{[OIII]\lambda 4363}$ emission line detection in the VIMOS-IFU data }
\label{4363 detection}
We aim to estimate $T_{e}-$based metallicities, which make use of the $[\ion{O}{iii}]\lambda 4363$ emission line to estimate the electron temperature, $T_{e}$ (\citealt{peimbert67}, \citealt{aller84}, \citealt{izotov06}). However, by integrating all the VIMOS-IFU fibres, the resulting spectra do not show the auroral line with $S/N>3$. We define the signal (S) as the flux value at the peak of the $[\ion{O}{iii}]\lambda 4363$ auroral line, and noise (N) is defined as the standard deviation evaluated in the continuum inside the spectral windows $4355-4360\AA$ and $4365-4370\AA$. Therefore, we should perform an additional treatment to be able to measure the auroral line with the desired S/N and, thus, get a reliable $T_{e}-$based oxygen abundance. 

We applied the following method: we first assume that $[\ion{O}{iii}]\lambda 5007$ and $[\ion{O}{iii}]\lambda 4363$ should have a similar spatial distribution across the \ion{H}{ii} region, because those are different emission lines from the same ionisation state of the oxygen, so we applied a criterion in the behaviour of the $[\ion{O}{iii}]\lambda 5007$ emission line. We defined the "jump" as the ratio between the flux value at the peak of the $[\ion{O}{iii}]\lambda 5007$ emission line and the semiquartile range of the spectral noise in the wavelength region of 4970-5040$\AA$. We choose the semiquartile range of spectral noise because it gives a representation of the flux level that is not sensitive to outliers (i.e. emission lines are outliers with respect to the flux noise). With this criterion, we can select those fibres with jumps high enough to reproduce an integrated spectrum where the $[\ion{O}{iii}]\lambda 4363$ emission line is detectable with S/N>3. 

\begin{figure}
   \centering
    \includegraphics[width=\hsize]{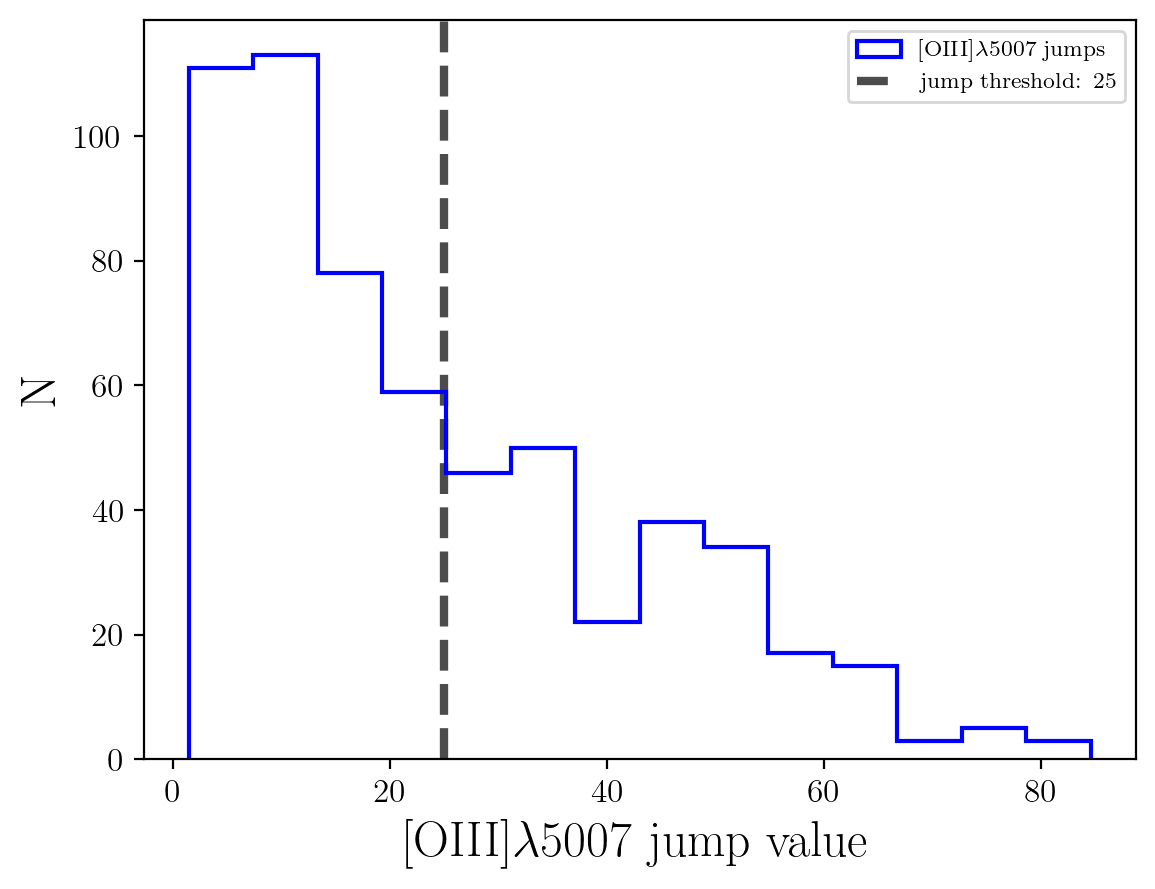}
      \caption{Distribution of jumps for SagDIG \ion{H}{ii} region spectral fibres shown in blue. The vertical \textcolor{black}{black dashed} line is the lower limit (25) to select those fibres where the integrated spectra of SagDIG \ion{H}{ii} show the auroral line with S/N>3. }
         \label{fig1}
\end{figure}

\begin{figure*}
   \centering
    \includegraphics[width=\hsize]{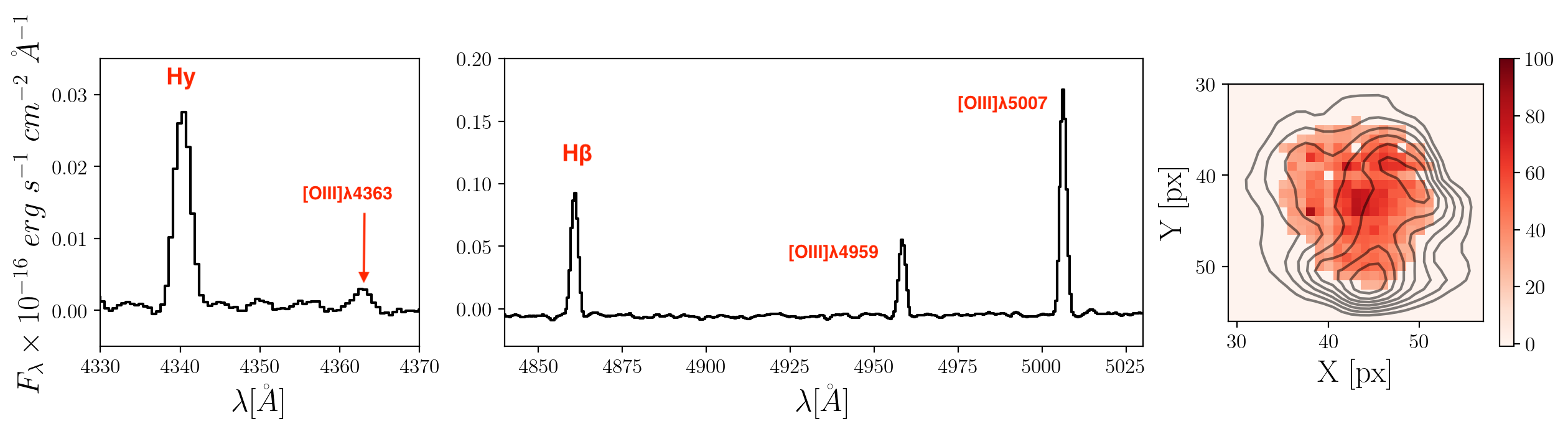}
      \caption{Integrated spectrum of the SagDIG \ion{H}{ii} region. The left panel shows the window wavelength of H$\gamma$ and $[\ion{O}{iii}]\lambda 4363$ emission lines. The middle panel shows the H$\beta$ and $[\ion{O}{iii}]\lambda\lambda 4959,5007$ emission lines. \textcolor{black}{The right panel shows the spatial distribution of selected fibres to generate the integrated spectrum of the SagDIG \ion{H}{ii} region with auroral detection. The colour code represents the jump value of each selected fibre. The grey contours represent the H$\beta$ emission of the nebula as reference.}}
         \label{fig2}
\end{figure*}

The distribution of jumps is shown in Figure \ref{fig1}, covering values between 0 and 90. To obtain the integrated spectrum where the auroral line is detectable, we employed an iterative procedure in the jump values that is described as follows: (i) we select those fibres based on a threshold jump value, starting in jump=1, then (ii) we integrate the spectrum of the selected fibres where the jump is greater than the threshold, and (iii) we evaluate the S/N of the auroral line in the integrated spectrum. If the estimated S/N of the auroral line is less than 3, we increase the value of the jump threshold and repeat the procedure. We found that the value of the threshold jump which shows the $[\ion{O}{iii}]\lambda 4363$ emission line with S/N$>3$ is jump=25, as the result of the integration of 234 fibres. \textcolor{black}{The threshold jump value that reproduces the integrated SagDIG \ion{H}{ii} region spectrum with a detectable S/N$>3$ is shown as the vertical black dashed line in Figure \ref{fig1}}.

\textcolor{black}{The integrated spectrum representing the SagDIG \ion{H}{ii} region is shown in the left and middle panels of Figure \ref{fig2}, where the auroral $[\ion{O}{iii}]\lambda 4363$ line is \textcolor{black}{marked} with the red arrow. The spatial distribution of the spectral fibres with jump>25 is shown in the right panel of Figure \ref{fig2}, where the colour code represents the jump value per spaxel. The H$\beta$ emission line map (grey contours) of the complete data cube is superimposed for reference.}

In our reduced and integrated spectrum, we do not detect underlying stellar absorption in the recombination H lines. Therefore, no additional subtraction was done after the data reduction. Single Gaussian functions were fitted to all available emission lines for the SagDIG integrated spectrum. To correct for internal absorption due to dust, we performed the Balmer decrement corrections. When is available, the procedure consists in using the theoretical H recombination emission line ratio $I_{H\alpha}/I_{H\beta} = 2.85$ (\citealt{hummer87}), adopting the Case B ($T_{e} \simeq 10^{4}K$). However, the $\mathrm{H\alpha}$ emission line falls outside the spectral coverage of the VIMOS-IFU observations used in this work. Therefore, following \citet{hummer87} we used the theoretical $I_{H\beta}/I_{H\gamma} = 2.137$ emission line ratio assuming the Case B. The correction is as follows: the reddening constant was calculated as $C_{H\gamma}= [\log(I_{H\beta}/I_{H\gamma})-\log(F_{H\beta}/F_{H\gamma})]/[f(H\beta)- f(H\gamma)]$. Where $f(\lambda) = \langle A(\lambda)/A(V) \rangle$ is the extinction law at a given wavelength. Following \citet{cardelli89}, we adopt $R_{V} = 3.1$. Finally, emission line measurements were corrected applying $I_{\lambda}/I_{H\gamma} = (F_{\lambda}/F_{H\gamma}) \times 10^{C_{H\gamma}[f(\lambda)-f(H\gamma)]}$. The emission line flux measurements are shown in Table \ref{table1}.

\begin{table}
\caption{Emission line flux measurements on the SagDIG \ion{H}{ii} region integrated spectrum. The first and second columns show the emission lines at their respective rest frame wavelength. The third column shows the flux measurements, and the fourth column shows the reddening corrected emission line measurements. The last row shows the SagDIG extinction coefficient C(H$\gamma$).}
\begin{tabular}{cccc}
\hline\hline
Ion& $\lambda \ [\AA]$ & $F_{\lambda}/F_{H\gamma}$ & $I_{\lambda}/I_{H\gamma}$ \\
\hline
$H_{\gamma}$         & 4340                    & 1.000 $\pm$ 0.001           & 1.000 $\pm$ 0.001         \\
$[\ion{O}{iii}]$    & 4363                    & 0.158 $\pm$ 0.011           & 0.127 $\pm$ 0.009        \\
$H_{\beta}$          & 4861                    & 3.534 $\pm$ 0.016           & 2.137 $\pm$ 0.001        \\
$[\ion{O}{iii}]$    & 4959                    & 2.002 $\pm$ 0.014           & 1.064 $\pm$ 0.079        \\
$[\ion{O}{iii}]$    & 5007                    & 5.820 $\pm$ 0.075           & 3.116 $\pm$ 0.167        \\
\hline
C(H$\gamma$)         & 0.293 $\pm$ 0.014 \\
\hline
\end{tabular}
\label{table1}
\end{table}

\subsection{EFOSC2 data}
\label{efosc2_data}
In order to estimate the total oxygen abundance, and explore variations of different ionisation states of oxygen across the SagDIG \ion{H}{ii} region, it is also needed the $[\ion{O}{ii}]\lambda3727$ emission line measurement. For this reason, we used the archival long-slit data coming from the ESO Faint Object Spectrograph Camera (v.2; EFOSC2, Buzzoni et al. 1984), which are also presented in \citet{saviane02}. The observations were taken in August 2000 (PI: Y. Momany) under the program 63.N-0726(B), at the ESO observatory in La Silla, Chile. EFOSC2 is a versatile instrument for low-resolution spectroscopy and imaging, and it is attached to the ESO 3.6m telescope at the dates of the observations. The EFOSC2 observations consist in $4\times1200$ s exposures by using the grism $\#7$ (327-5250 nm, and $0.96\AA \ $px$^{-1}$ dispersion), and the grism $\#9$  (470-677 nm, and $1\AA \ $px$^{-1}$ dispersion), both with a $1.5''$ slit.

\begin{figure}
   \centering
    \includegraphics[width=\hsize]{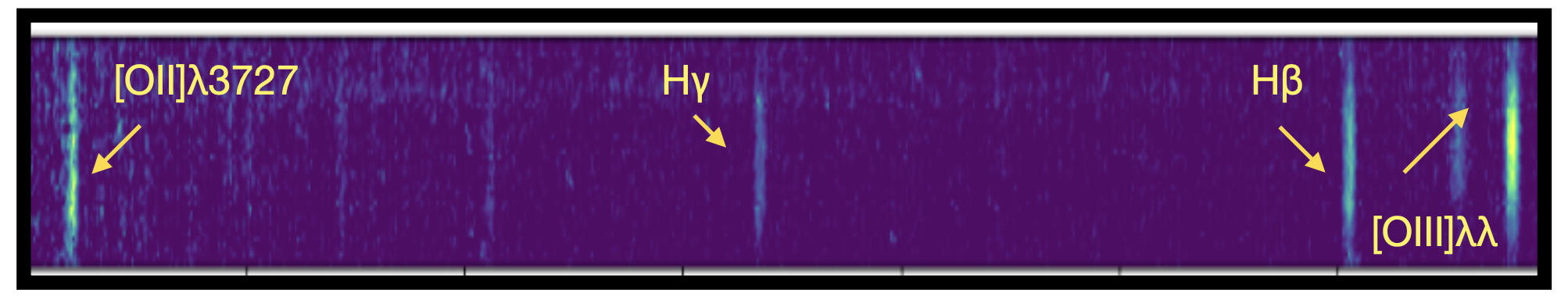}
      \caption{EFOSC2 long-slit flux calibrated two-dimensional frame of the SagDIG \ion{H}{ii} region. Yellow arrows show the detected emission lines in each slit column.}
         \label{fig3}
\end{figure}

Each scientific frame was reduced independently with IRAF\footnote{IRAF is distributed by the National Optical Astronomy Observatories, which are operated by the Association of Universities for Research in Astronomy, Inc., under a cooperative agreement with the National Science Foundation.} We performed bias subtraction, flat normalisation, wavelength calibration, and flux calibration in each \textcolor{black}{position along the slit} which covers the \ion{H}{ii} region. We select 10 \textcolor{black}{slit pixels} outside the \ion{H}{ii} region to extract the sky spectrum and stellar continuum, and apply the subtraction to each slit pixel of the \ion{H}{ii} region. The flux calibration was done by using a spectrophotometric standard star from \citeauthor{hamuy92} (\citeyear{hamuy92}, \citeyear{hamuy94}): The G-type LTT-1020 ($V = 11.52, \ B-V =0.56 $). The final products are flux-calibrated in units of $10^{-16}\mathrm{erg\ s^{-1}\ cm^{-2}\ \AA^{-1}}$. The dust reddening correction was applied in the same way as the VIMOS-IFU data. The flux-calibrated two-dimensional frame is shown in Figure \ref{fig3}, with the detected emission lines shown in yellow.

\begin{figure*}
\centering
\includegraphics[width=\hsize]{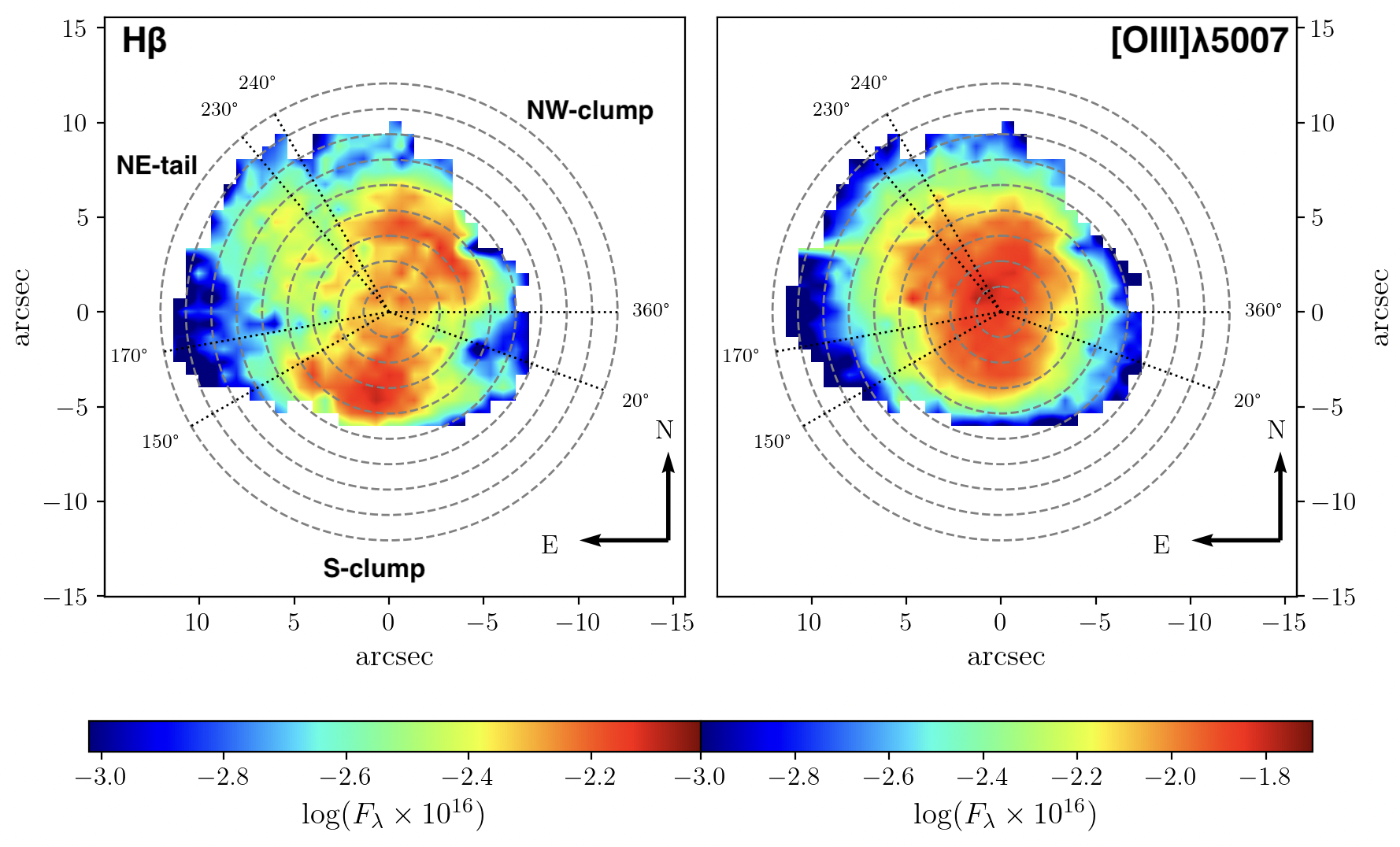}
\caption{The SagDIG \ion{H}{ii} region emission line maps: left panel shows the H$\beta$ emission line distribution. The right panel shows the $[\ion{O}{iii}]\lambda 5007$ emission line distribution. The colour map represents the flux on each emission line, and gray dashed lines are circles of increasing radius of $1.34''\ (2$ px) units up to $12.06''\ (18$ px). Black dotted lines show the angles that separate the S-clump, the NE-tail, and the NW-clump.}
\label{fig4}
\end{figure*}

%--------------------------------------------------------------------
\section{Data analysis}
\subsection{Spatially-resolved flux distributions}
\label{emission_line_maps}

Once the VIMOS-IFU fibre spectra were reduced, we proceeded to study the gas-phase structures of the SagDIG nebula. The flux distributions presented in this subsection were made by measuring the flux of the detectable (S/N$>3$) strong emission lines H$\beta$ and $[\ion{O}{iii}]\lambda 5007$, \textcolor{black}{fitting Gaussian curves on each continuum and sky subtracted spectrum for all spectral fibres of the VIMOS cube.}

Figure \ref{fig4} shows the SagDIG \ion{H}{ii} region spatial flux distribution for H$\beta$ in the left panel, and $[\ion{O}{iii}]\lambda 5007$ in the right panel. Gray dashed circles of increasing radius of $1.34''\ (2$ px) units up to $12.06''\ (18$ px) are also shown for a posterior analysis of the flux density profiles, centred at $\alpha= 19^{h}30^{m}3.2^{s}$, $\delta=-14\degree 41' 29.94''$ (J2000). The centre was settled as the intermediate point between the two H$\beta$ clumps. Additionally, we applied a bi-linear interpolation in the flux maps to smooth out spatial variations between the pixels and to remove artefacts.\\ 

\begin{figure}
\centering
\includegraphics[width=6cm]{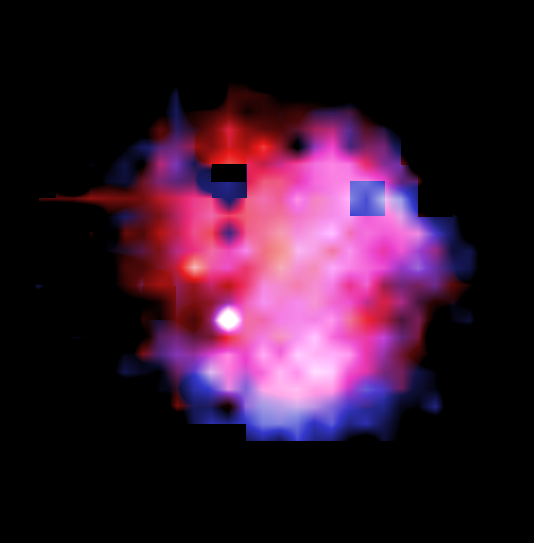}
\caption{Composite image of the flux distribution shown in Figure \ref{fig4}. The blue colour represent the H$\beta$ flux distribution, whereas the red colour represents the $[\ion{O}{iii}]\lambda 5007$ flux distribution.}
\label{fig5aaa}
\end{figure}

The H$\beta$ flux distribution reveals that the SagDIG \ion{H}{ii} region has two main clumps, one in the southern region and the other one in the northwest direction, both at $\sim 5''$ from the centre. It is also noted in the left panel that there is a small tail-like structure from the centre towards the northeast up to $\sim8''$. \\
\\
The right panel of Figure \ref{fig4} shows the $[\ion{O}{iii}]\lambda 5007$ flux distribution which seems to cover a greater area in the map than the H$\beta$ emission, which is indicated with reddish colours up to $\sim8''$ from the centre. This is clearer with when the composite image of the left and right panels of Figure \ref{fig4} is created, as shown in Figure \ref{fig5aaa}, with blueish and reddish colors for the H$\beta$ and $[\ion{O}{iii}]\lambda 5007$ emission, respectively.

Radial flux density profiles were generated to study the H$\beta$ and $[\ion{O}{iii}]\lambda 5007$ structures, defined as $\Sigma_{\lambda} = \sum F_{\lambda}/A_{ring}$, where $\sum F_{\lambda}$ is the sum of the flux of a given emission line of the fibres inside the area between two consecutive concentric circles of radius $r_{in}$ and $r_{out}$ described as 

\begin{equation}
A_{ring} = \int_{\phi_{0}}^{\phi_{f}}\int_{r_{in}}^{r_{out}}rdrd\phi =  \frac{1}{2} (\phi_{f} - \phi_{0})(r_{out}^{2}-r_{in}^{2}) 
\end{equation}

shown in Figure \ref{fig4}. $\phi_{0}$ and $\phi_{f}$ are the angles that contain each structure (clockwise in the x-axis from the centre of the concentric circles, from left to right), i.e., the south clump, the north-west clump, and the north-east tail-like structure. 

To generate the south clump (hereafter S-clump) flux density profile, we select all fibres between $20\degree<\phi<150\degree$. The north-east tail-like structure (hereafter NE-tail) radial profile was produced by selecting fibres between $170\degree<\phi<230\degree$. Finally, the north-west clump (hereafter NW-clump) radial profile was produced by using the fibres between $240\degree<\phi<360\degree$. The angular selection of the H$\beta$ structures is shown with the black dotted lines in both panels of Figure \ref{fig4}. The flux density profiles have units of $\mathrm{10^{-16}\  ergs\ s^{-1}\ cm^{-2}\ \AA^{-1}\ \mathrm{arcsec}^{-2}}$.

\begin{figure}
\centering
\includegraphics[width=\hsize]{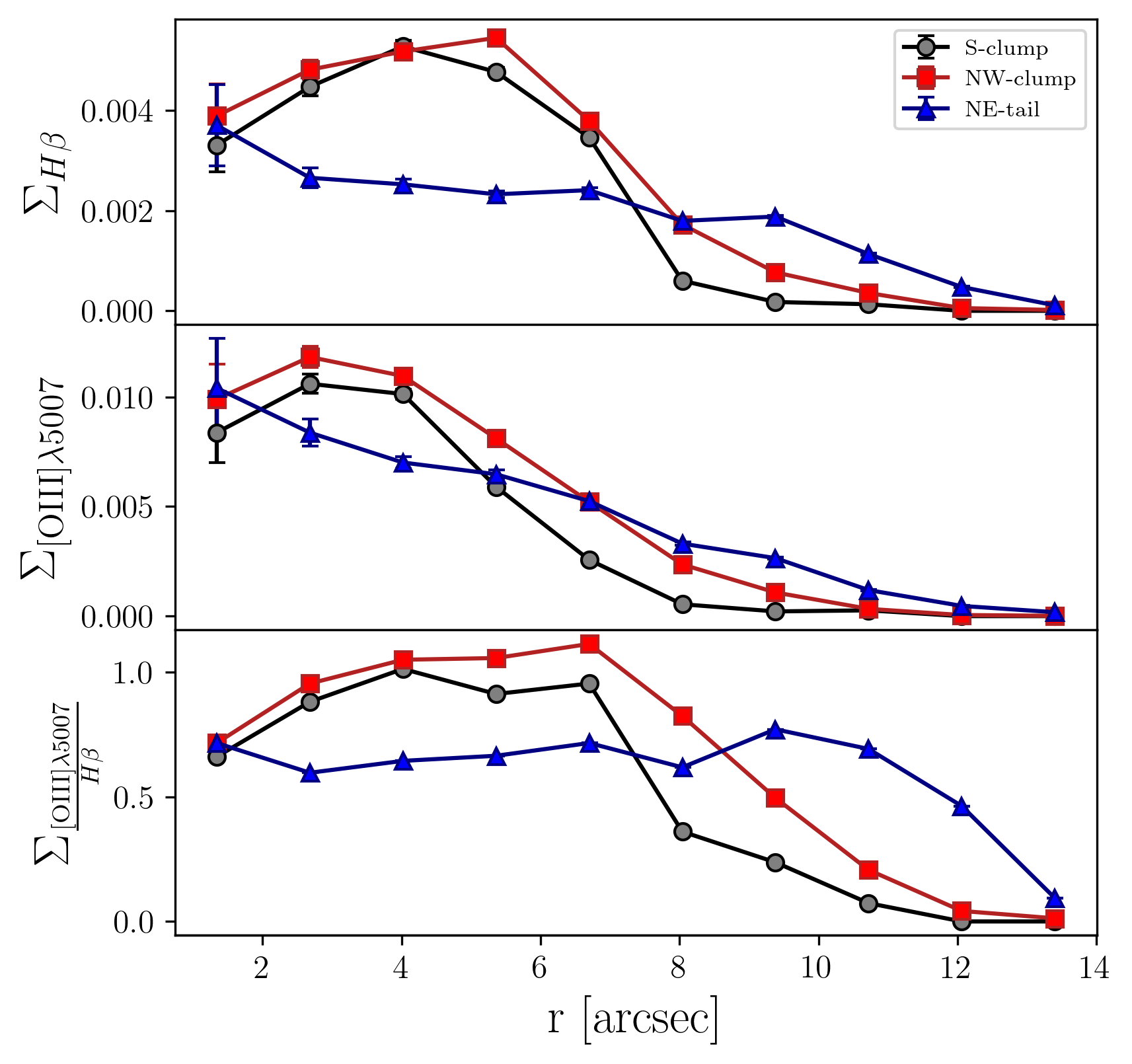}
\caption{Flux density profiles for the S-clump, NW-clump, and NE-tail structures present in the SagDIG nebula, with black \textcolor{black}{dots}, red \textcolor{black}{squares}, and blue \textcolor{black}{triangles}, respectively. The top panel shows the H$\beta$ flux density profiles. The middle panel shows the $[\ion{O}{iii}]\lambda 5007$ flux density profiles. \textcolor{black}{The bottom panel shows the $[\ion{O}{iii}]\lambda 5007$ / H$\beta$ flux density profile.}}
\label{fig5}
\end{figure}

The top and \textcolor{black}{middle} panels of Figure \ref{fig5} show the H$\beta$, and $[\ion{O}{iii}]\lambda 5007$ radial flux density profiles, respectively, for the three structures. Black lines represent the S-clump, red lines represent the NW-clump, and blue lines represent the NE-tail. 
\\
\\
The upper panel of Figure \ref{fig5} indicates that both the S-clump and the NW-clump are similar in terms of H$\beta$ flux density. Both increases in flux density up to reach a peak between $\sim$4$''-6''$, then decrease in flux density up to $\sim$8$''$ and $\sim$9$''$ for the S-clump and NW-clump, respectively.

Regarding the $[\ion{O}{iii}]\lambda 5007$ flux density both clumps show similar behaviour. However, the $[\ion{O}{iii}]\lambda 5007$ flux density is almost twice as seen in H$\beta$ flux density. Both peaks at $\sim$3$''$. Comparing the H$\beta$ and the $[\ion{O}{iii}]\lambda 5007$ flux densities, the latter is more concentrated in the centre than the former. On the other hand, the NE-tail shows a different flux density distribution in H$\beta$ and lower flux-density with respect to the clumps. However, despite $[\ion{O}{iii}]\lambda 5007$ flux density profile of the NE-tail show a gradual radial decrement, the profile is similar with respect to the clumps, suggesting that the distribution of oxygen is more symmetric in the \ion{H}{ii} region compared with the hydrogen flux distribution.\\ 

\textcolor{black}{The bottom panel of Figure \ref{fig5} presents the flux density profile of the emission line ratio $[\ion{O}{iii}]\lambda 5007$/H$\beta$ for the three substructures. Both clump's flux-density profiles reveal similar trends: the S-clump shows values closer to 1, while the NW-clump presents values slightly higher than 1. Then, both flux-density profiles decrease their emission line ratio for radii greater than 6$''$. This feature suggests that $\mathrm{O^{++}}$ is more centrally concentrated with respect to $\mathrm{H^{+}}$. Otherwise, the flux density profile of this emission line ratio should show a flat behaviour.}

The H$\beta$ and the $[\ion{O}{iii}]\lambda 5007$ emission maps and flux density profiles are showing signs of the so-called "stratification" in \ion{H}{ii} regions (\citealt{osterbrock06}): SNe explosions and UV radiation from massive stars (OB spectral type) located in \ion{H}{ii} regions can inject a huge energy budget to their surrounding gas. Because radiation density decrease with distance, the central regions host highly ionised species (such as $\mathrm{O^{++}}$, ), whereas the outskirts present lower ionisation species (such as $\mathrm{H^{+}}$, $\mathrm{O^{+}}$, $\mathrm{N^{+}}$, and $\mathrm{S^{+}}$), where the $\mathrm{H^{+}}$ prevails over $\mathrm{O^{++}}$. This creates a stratified ionisation structure with different iosized zones for different atoms, as is seen in some galactic \ion{H}{ii} regions and the Large Magellanic Cloud (\citealt{sanchez13}, \citealt{barman22}, \citealt{kreckel24}).

\subsection{The search for stratification in the EFOSC long-slit data}
\label{efosc_spatial_variations}
\begin{figure}
\centering
\includegraphics[width=\hsize]{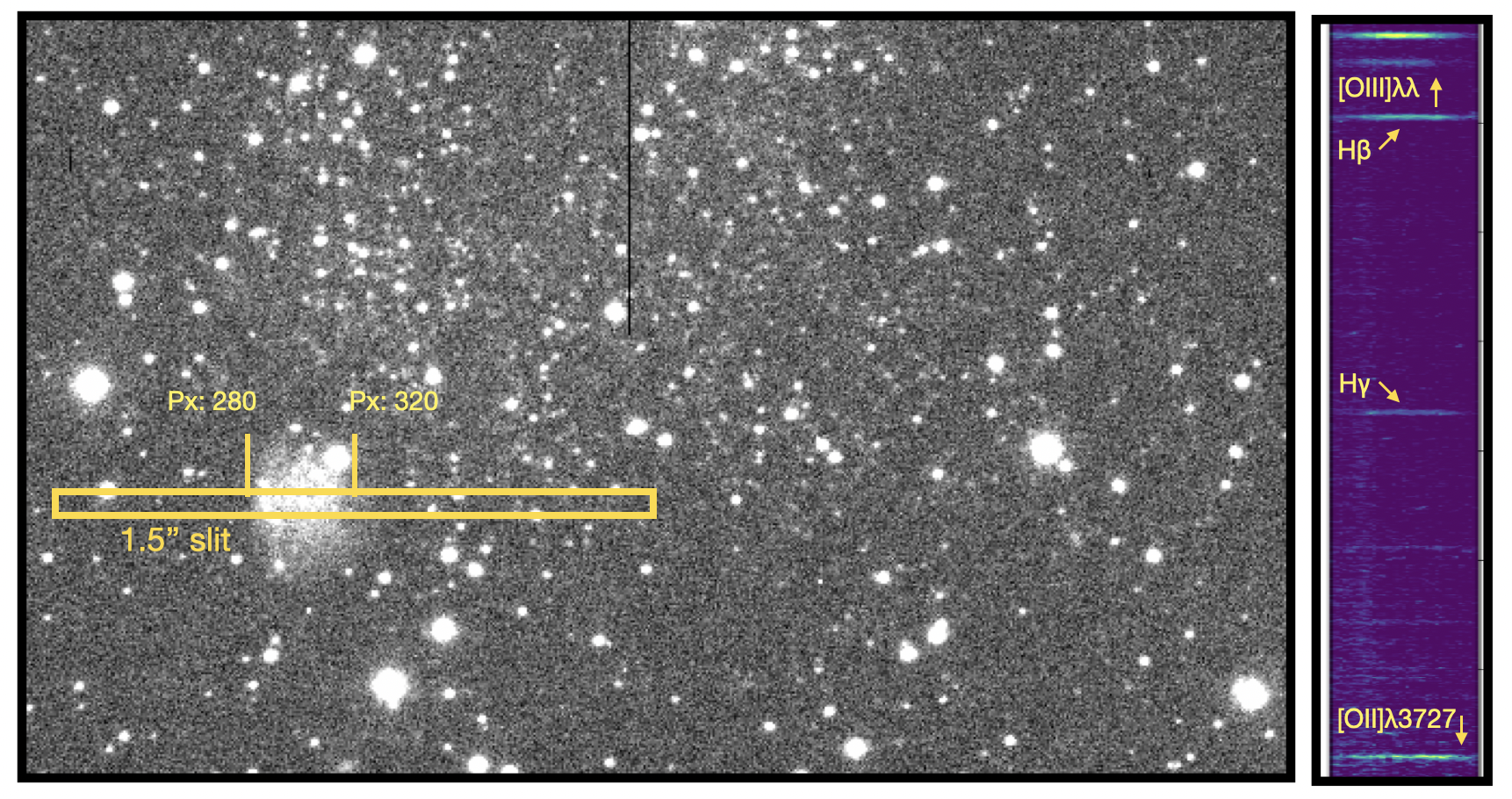}
\caption{Left panel: EFOSC2 pre-imaging frame of the SagDIG \ion{H}{ii} region. The yellow rectangle shows the slit position settled to extract the two-dimensional frame spectra. Vertical yellow lines show the area covered by \ion{H}{ii} region, which is from column 280 to column 320. Right panel: two-dimensional frame spectra, where the emission lines detected are shown with yellow arrows.}
\label{fig6}
\end{figure}

\begin{figure}
\centering
\includegraphics[width=\hsize]{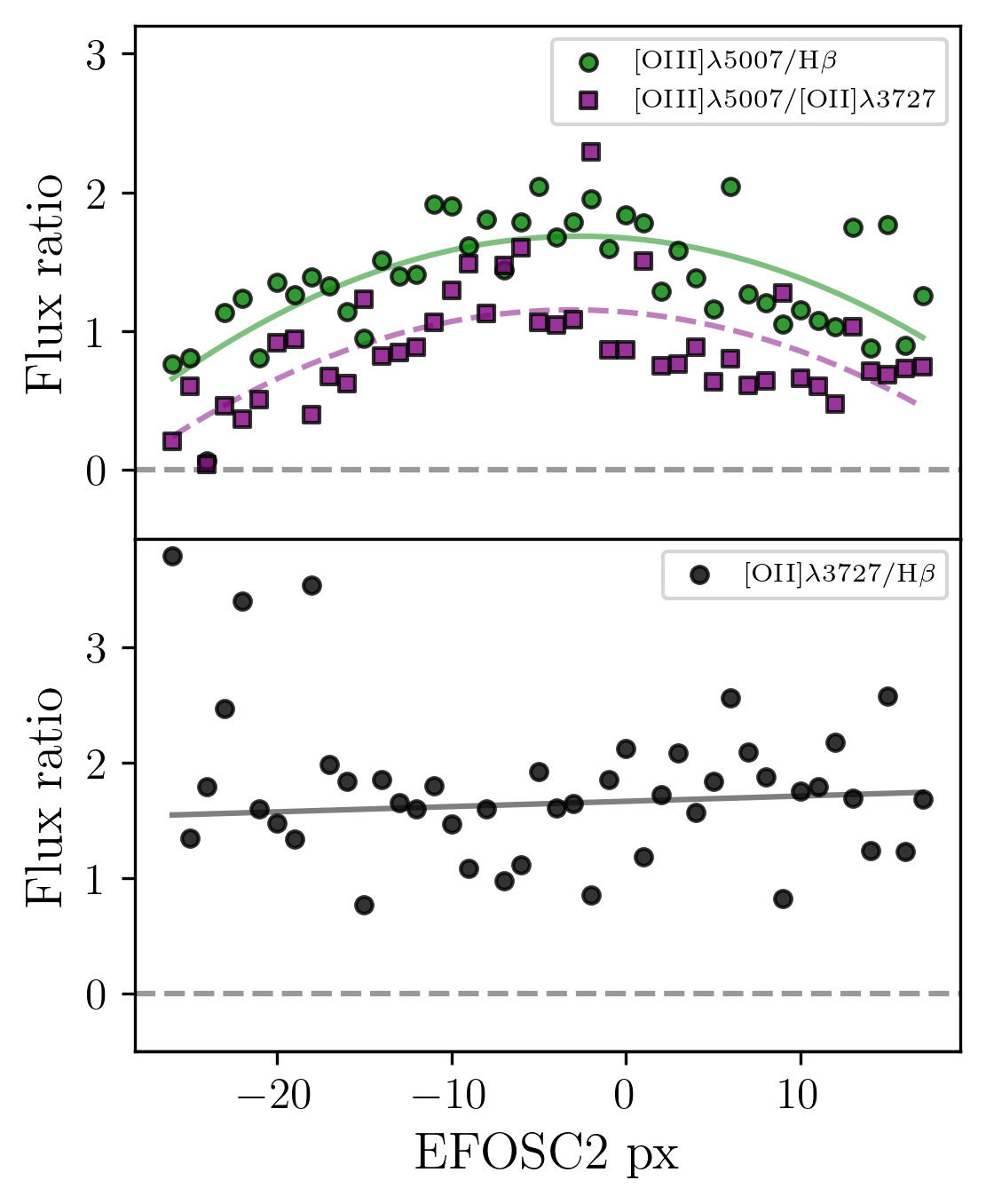}
\caption{Emission line ratios across the slit column, which represent the East-West direction coordinate in the SagDIG \ion{H}{ii} region. The upper panel shows the $[\ion{O}{iii}]\lambda 5007 / \mathrm{H}\beta$ and the $[\ion{O}{iii}]\lambda 5007 / [\ion{O}{ii}]\lambda 3727$ emission line ratios, \textcolor{black}{with green dots and purple squares, respectively}. The lower panel shows the $[\ion{O}{ii}]\lambda 3727 / \mathrm{H}\beta$ emission line ratio with the black dots.}
\label{fig7}
\end{figure}
In this section we used the EFOSC2 long-slit spectrum to further study the ionization stratification of the \ion{H}{ii} region using a more direct approach that compare two oxygen ionization levels ($\mathrm{O^{+}}$ and $\mathrm{O^{++ }}$). The slit position in the SagDIG \ion{H}{ii} region is shown in the left image of Figure \ref{fig6}, which corresponds to the pre-imaging frame of the EFOSC2 long-slit observations.

The slit position is shown with the yellow rectangle, and the vertical yellow lines indicate the \textcolor{black}{position along the slit}  where the SagDIG \ion{H}{ii} region is covered (\textcolor{black}{$x=280$} to \textcolor{black}{$x=320$}). The right image of Figure \ref{fig6} shows the spectra of the SagDIG \ion{H}{ii} region inside the slit area from \textcolor{black}{pixel} 280 to 320 in the two-dimensional frame from left to right, respectively. As is seen in the right panel of Figure \ref{fig6}, H$\beta$, H$\gamma$, $[\ion{O}{ii}]\lambda 3727$, and $[\ion{O}{iii}]\lambda\lambda 4959,5007$ emission lines are detected.\\
\\
To confirm the stratification, it is needed to see if $\mathrm{O^{+}}$ is spatially distributed (across the slit region covered) in a similar way as $\mathrm{H^{+}}$, with respect to $\mathrm{O^{++}}$. Therefore, we proceed to study the emission line ratios $[\ion{O}{iii}]\lambda 5007 / H\beta$ and $[\ion{O}{iii}]\lambda 5007 / [\ion{O}{ii}]\lambda 3727$ as a function of the \textcolor{black}{position along the slit}. It is important to note that the \textcolor{black}{position along the slit} represents the spatial coordinates in the x-axis. Therefore, we are studying here the behaviour of the emission line ratios in East-West direction \textcolor{black}{along the slit} area that covers the \ion{H}{ii} region. \\

\textcolor{black}{Figure \ref{fig7} presents the distribution of the emission line ratios between H$\beta$, $[\ion{O}{ii}]\lambda3727$, and $[\ion{O}{iii}]\lambda5007$ along the slit}. The x-axis represents the position along the slit, where 0 is the centre of the \ion{H}{ii} region settled in Figure \ref{fig4}. The upper panel of Figure \ref{fig7} presents the emission line ratios $[\ion{O}{iii}]\lambda 5007 / \mathrm{H}\beta$ and the $[\ion{O}{iii}]\lambda 5007 / [\ion{O}{ii}]\lambda 3727$, with green dots and purple squares, respectively, together with a 2nd degree polynomial fitting for each emission line following the same colours. This panel reveals that both emission line ratios behave similarly, as high values of the emission line ratios in the centre with respect to the outskirts. The lower panel presents the emission line ratio $[\ion{O}{ii}]\lambda 3727 / \mathrm{H}\beta$ with black dots, and their respective linear fitting with the grey line, revealing that the emission line is behaving flat across the position along the slit at $[\ion{O}{ii}]\lambda 3727 / \mathrm{H}\beta$ $\sim 1.6$. The linear fitting suggests a slope $0.004\pm 0.006$.

Those panel indicates that (i) the $\mathrm{O^{++}}$ is more concentrated in the centre of the \ion{H}{ii} region in comparison to both the $\mathrm{O^{+}}$ and the $\mathrm{H^{+}}$, and (ii) the $\mathrm{O^{+}}$ is distributed in a similar way as the $\mathrm{H^{+}}$ \textcolor{black}{along} the slit area that covers the \ion{H}{ii} region. These features are giving the same information as the emission line maps and the flux-density profiles shown in Figure \ref{fig5} and Figure \ref{fig6}. This validates, thus, that the SagDIG \ion{H}{ii} region is a dynamic nebula.

\subsection{$T_{e}-$based oxygen abundances}
\label{metallicities}

\begin{table*}
\label{table2}
\caption{Electron temperature and ionic oxygen abundances estimate for SagDIG. The second column shows the $T_{e}$ estimates, the third and fourth columns shows the ionic oxygen abundances $\mathrm{O^{+}/H}$ and $\mathrm{O^{++}/H}$, respectively. The fifth and sixth columns shows the total oxygen abundances with the $T_{e}$, and the $R_{23}$ empirical calibrator, respectively.}
\centering
\begin{tabular}{ccccc}
\hline \hline
                                    & $T_{e}$ & $12+\mathrm{log(O^{+}/H)}$ & $12+\mathrm{log(O^{++}/H)}$ & $12+\mathrm{log(O/H)}$ \\ \hline
VIMOS mock slit - Pyneb             & $22165 \pm 1122$K&   $6.98\pm 0.02$& $ 6.83\pm 0.06$ & $7.23\pm 0.04$\\
VIMOS mock slit - I06             &  --                &   $6.47\pm 0.03$& $ 6.83\pm 0.03$ & $7.00\pm 0.03$\\
Central region - Figure \ref{fig9bbb} & $23390\pm1094$K & $7.06 \pm 0.06$ & $6.78 \pm 0.04$ & $7.24\pm0.05$  \\ 
External region - Figure \ref{fig9bbb} & $21941\pm2572$K & $7.45 \pm 0.15$ & $6.53 \pm 0.12$ & $7.50\pm0.14$ \\
\hline

\hline

\end{tabular}
\end{table*}
We aim to estimate oxygen abundances with the so-called "direct method" (\citealt{peimbert67}, \citealt{aller84}, \citealt{izotov06}), so the integrated spectrum of the SagDIG \ion{H}{ii} region is used to estimate metallicities. This method makes use of emission line ratios sensitive to the electron temperature, $T_{e}$, together with the computation of the respective line emissivities. The estimated values are presented in Table 2 and are discussed in Appendix \ref{ap1:metallicities}.\\
\\
The $T_{e}-$sensitive emission line ratios are estimated by using the weak auroral lines such as $[\ion{O}{iii}]\lambda 4363$ or $[\ion{N}{ii}]\lambda 5755$. Measuring the auroral line in each spectrum fibre is quite difficult because this emission line is $\sim3-4$ order of magnitudes fainter than the Balmer emission lines (\citealt{maiolinomanucci19}), and can get lost in the spectral noise. The use of the integrated spectra reduces the spectral noise in comparison with the individual fibres's spectra. As is seen in Figure \ref{fig2}, the integrated spectrum of SagDIG shows the auroral $[\ion{O}{iii}]\lambda 4363$ emission line with $\mathrm{S/N>3}$.

This method needs also the electron density, $N_{e}$, to compute the oxygen abundances. However, our data does not cover the wavelength range for the estimation of $N_{e}$ through emission lines such as $[\ion{S}{ii}]\lambda\lambda 6717,6731$. So we assume $N_{e}=100$ cm$^{-3}$ according to the low-density limit (\citealt{hummer87}).\\

We used the $[\ion{O}{iii}](\lambda4959+\lambda5007)/\lambda4363$ emission line ratio to estimate $T_{e}\mathrm{([\ion{O}{iii}])}$. We did this by using the \textit{getTemDen} module of the Pyneb python package (\citealt{luridiana15}). Then, $T_{e}\mathrm{([\ion{O}{ii}])}$ was calculated by using the linear relation $T_{e}\mathrm{([\ion{O}{ii}])} = 0.7\times T_{e}\mathrm{([\ion{O}{iii}])}+3000$ calibrated by \citet{campbell86} from the \citet{stasinska82} photoionisation models.

The $T_{e}-$based oxygen abundance was estimated with two different approaches: (i) with the \textit{getIonAbundance} Pyneb module which uses Eq. 4 of \citet{luridiana15}, and (ii) using Eq. 3 and Eq. 5 from \citeauthor{izotov06} (\citeyear{izotov06}, hereafter I06), which were estimated by using photoionisation grid models of \ion{H}{ii} regions. The use of Eq. 3 and Eq. 5 of I06 requires $T_{e}$ to estimate the ionic oxygen abundances $\mathrm{O^{+}/H}$ and $\mathrm{O^{++}/H}$. So we used the estimated values of the \textit{getTemDen} Pyneb module for $T_{e}$. The $T_{e}$ estimates are shown in Table 2. \\
\\
\textcolor{black}{We derived $T_{e} = 17683\pm1254$ considering the integrated spectrum of the SagDIG \ion{H}{ii} region (Table \ref{table1} and Figure \ref{fig2}). The total oxygen abundance is defined as $12+\mathrm{log(O/H)}$, where $\mathrm{O/H = (O^{+}/H^{+} + O^{++}/H^{+})}$. The VIMOS-IFU spectral range does not cover the wavelength region where $[\ion{O}{ii}]\lambda 3727$ is found. For this reason, we used the $[\ion{O}{ii}]\lambda 3727$ flux measurement from S02 to estimate the $\mathrm{O^{+}/H}$ ionic abundance. However, that measurement comes from a $1.5''$ EFOSC2 long-slit spectrum, covering the central part of the \ion{H}{ii} region, as shown in Figure \ref{fig6}. The combination of the VIMOS-IFU integrated spectrum and the EFOSC2 long-slit spectrum can introduce both instrumental biases and aperture effects}. \textcolor{black}{For this reason, we reproduce a mock-slit in the VIMOS IFU data following the EFOSC2 long-slit observations of S02. Our mock-slit is 2px wide, i.e., $1.34^{\prime\prime}$ and has the same position and orientation of S02.}\\
\\

\textcolor{black}{While the mock-slit aperture is slighly narrower than the actual slit employed in S02 and in spite of likely imperfect centring, we found that the EFOSC2 and the mock-slit measurements return ionic $\mathrm{O^{++}/H}$ abundances in excellent agreement. Therefore, we decided to proceed and derive a tentative $T_{e}-$based metallicity by combining the EFOSC2 measurements of $[\ion{O}{ii}]\lambda3727$ emission line with the $T_{e}$ obtained from the VIMOS mock-slit $[\ion{O}{iii}]\lambda4363$ emission line plus the lines in common between the two datasets. The complete procedure on the different approaches to estimate $T_{e}$ and oxygen abundances is described in Appendix \ref{ap1:metallicities}}.\\
\\
\textcolor{black}{We derived $T_{e}[\ion{O}{iii}]= 22165\pm 1122$ K, $T_{e}[\ion{O}{ii}]= 18515\pm 987$ K, and $T_{e}-$based total oxygen abundance of $\mathrm{12+log(O/H)}= 7.23\pm0.04$ dex using Pyneb. However, I06 calibration results in $\mathrm{12+log(O/H)}= 7.00\pm0.03$ dex. The discrepancy between Pyneb and I06 $T_{e}-$based oxygen abundances have been previously explored by \citet{laseter24} in a sample of high-redshift galaxies, showing differences ($\sim0.1$ dex on average and 0.2 dex in extreme cases) consistent with our results. Further investigation led us to conclude that the difference in metallicities is produced by the estimated $\mathrm{O^{+}/H}$, as is shown in Table 2, probably because I06 brings an empirical relation between the flux $[\ion{O}{ii}]\lambda 3727$ and $T_{e}[\ion{O}{iii}]$ with the ionic abundance $\mathrm{O^{+}/H}$. Therefore, we proceed to use the Pyneb $T_{e}-$based oxygen abundances for the interpretation of the results presented in Section \ref{results}.} \\
\\
\textcolor{black}{The difference between $T_{e}$ using the integrated spectrum and the mock-slit suggest that there are temperature fluctuations present in the SagDIG \ion{H}{ii} region with subsequent aperture effects.}

\subsection{{Intrinsic biases of the $T_{e}-$sensitive method.}}  
\label{Te-biases}
Several works in the literature have discussed the well-known biases affecting the direct method (\citealt{kobulnicky96}, \citealt{pilyugin16}, \citealt{cameron23}), such as (i) temperature fluctuations, which in general leads to overestimated temperatures and subsequent underestimated metallicities, because of the dependence of the emissivities with the statistical factor $e^{-h\nu/kT_{e}}$ (\citealt{aller84}, \citealt{osterbrock06}, \citealt{maiolinomanucci19}. (ii) Different ways to estimate the electron temperature, such as the use of emission line ratios, and the use of the Balmer or Paschen jumps (\citealt{peimbert67}). (iii) $\mathrm{O^{++}}$ abundances estimated from recombination lines can return metallicities up to three times those estimated from the collisional de-excitation lines (\citealt{esteban14}).

Despite this, the direct method is still the best choice to estimate oxygen abundances, because the so-called "empirical methods" (which make use of different strong emission lines depending on the choice of the calibrator) show oxygen abundance differences of up to 0.7 dex between them (\citealt{kewley08}, \citealt{poetrodjojo21}).

We are able to address the temperature fluctuation problem, since our integrated spectrum has a detectable $[\ion{O}{iii}]\lambda4363$ emission line, as follows: \citet{cameron23} used the \citet{katz22} dwarf galaxy ($M_{halo}=10^{10}M_{\odot}$, $M_{gas}=3.5\times10^{8}M_{\odot}$, and central metallicity of $0.1Z_{\odot}$) simulated with the RAMSES-RTZ code. They tackle the temperature fluctuations problem by introducing the "line temperature", $T_{line}$, which links the emissivities of the emission lines along a range of probed temperatures (see their Figure 1). With this approach, the authors report the relation $T_{line}=0.6T_{ratio}+3258\ K$, where $T_{ratio}$ is the standard $T_{e}$ estimated from the $\mathrm{[OIII]}$ ratio.\\
\\ 
\textcolor{black}{Applying the $T_{line}$ correction we obtain $12+\mathrm{log(O/H)_{corr}} = 7.50\pm 0.08$ ($T_{line}=16557\pm1093$) with our flux measurements. The correction leads to an increase in the oxygen abundances of $\sim$0.3 dex.}

\subsection{$T_{e}$ and oxygen abundance spatial variations in the SagDIG \ion{H}{ii} region}
\begin{figure}
\centering
\includegraphics[width=\hsize]{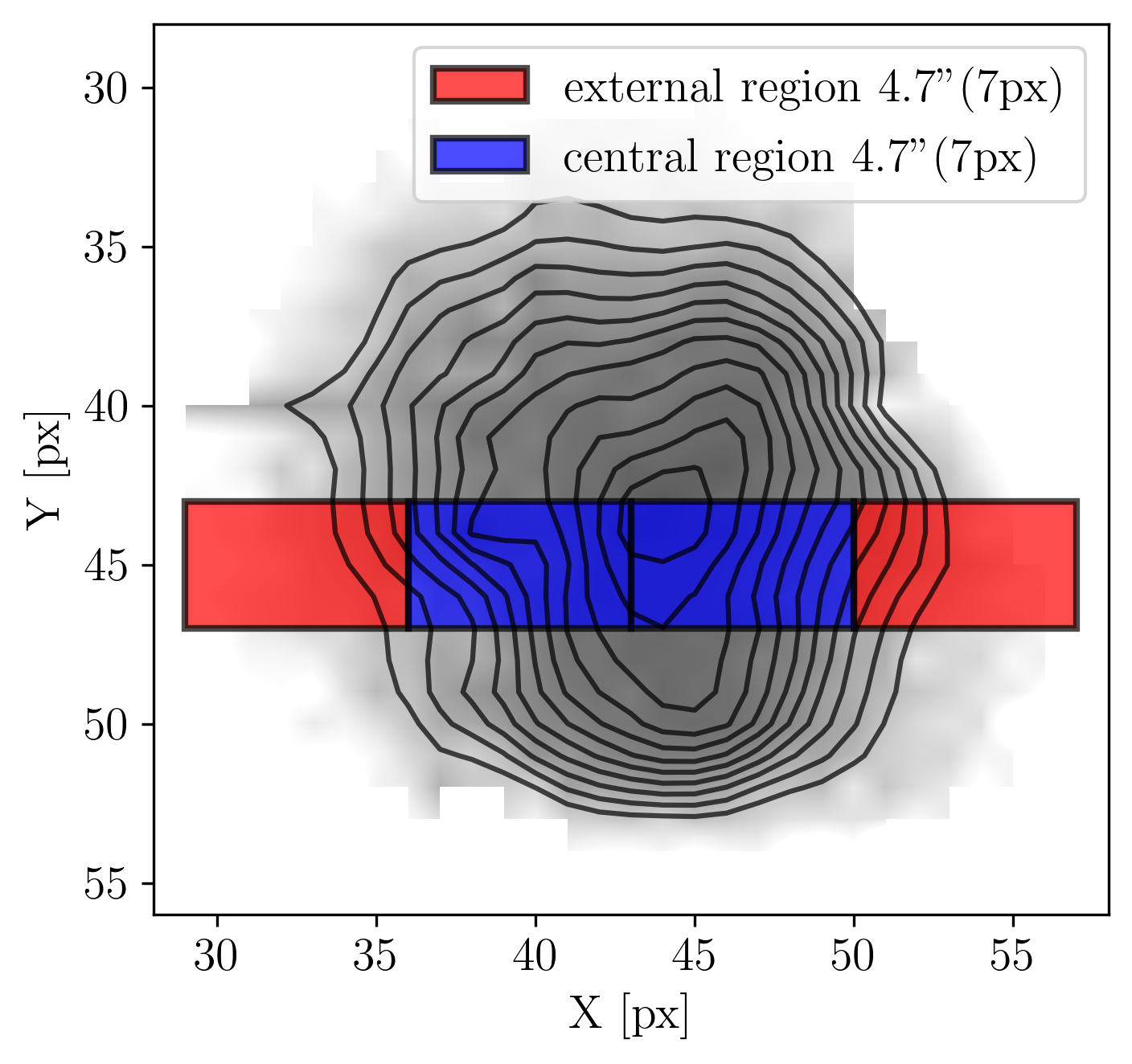}
\caption{Simulated slit in the IFU data following the EFOSC2 long-slit observations of S02, compared with the [\ion{O}{iii}]$\lambda5007$ emission line map, shown with the black contours. The simulated slit was divided in four sections of equal area, where first and fourth represents the external region with red, and the second and third represents the central region with blue, from left to right, respectively.}
\label{fig9bbb}
\end{figure}

In Section \ref{metallicities} we show that the $T_{e}$ of the \ion{H}{ii} region is $\sim$22100 K. However, it is possible to explore $T_{e}$ and metallicity variations inside the area covered by the EFOSC2 observations (coloured area in Figure \ref{fig9bbb}). This is because we flux-calibrated the two-dimensional frame of the S02 long-slit data (Section \ref{efosc2_data}).

To explore possible spatial variations, we (i) select a central and external region in the VIMOS-IFU simulated slit to estimate $T_{e}$ and the ionic abundance $\mathrm{O^{++}/H}$, and (ii) apply the same spatial cuts in the EFOSC2 two-dimensional frame (Figure \ref{fig9bbb}), to get the respective measure of the $[\ion{O}{ii}]\lambda3727$ emission line flux and a subsequent $\mathrm{O^{+}/H}$ ionic abundance, for a central and external region of the \ion{H}{ii} region.

We divided the simulated slit into 4 sections of the same area, where the external region is defined as the first and fourth sections, and the central region is defined as the second and third section from left to right, as is shown in Figure \ref{fig9bbb} with red and blue. The same was done in the two-dimensional frame of the EFOSC-long slit data. Each section in the simulated slit in the VIMOS-IFU data is 7px long, so both the central and external regions are covering 14px long, which is $9.4''$.\\
\\
\textcolor{black}{The central region has a $T_{e} = 23390 \pm 1096$ and $12+\mathrm{log(O/H)} = 7.24 \pm 0.05$, whereas the external region has a $T_{e}=21941\pm 2572$ and  $12+\mathrm{log(O/H)}=7.50 \pm 0.08$, showing a decrease of $\sim$1500 K and an increase of $\sim0.25$ dex towards the outskirts. }

\textcolor{black}{On the other hand, the individual ionic abundances for the central region and external are $12+\mathrm{log(O^{+}/H)} =7.06 \pm 0.06$ and $12+\mathrm{log(O^{++}/H)}=6.78 \pm 0.04$, and $12+\mathrm{log(O^{+}/H)} =7.45 \pm 0.15$ and $12+\mathrm{log(O^{++}/H)}=6.53 \pm 0.12$, respectively. }
Those results suggest that (i) the centre is slightly hotter and more metal-poor than the outskirts, (ii) the centre has more $\mathrm{O^{++}}$ ionic abundance than the outskirts, and (iii) the outskirts have more $\mathrm{O^{+}}$ ionic abundance than the centre, in line with the stratified composition of the SagDIG \ion{H}{ii} region observed in Sections \ref{emission_line_maps} and \ref{efosc_spatial_variations}.

%--------------------------------------------------------------------
\section{Results and Discussion}
\label{results}
\subsection{The role of star formation in the SagDIG \ion{H}{ii} region}
\label{SagDIG_photometry}

\begin{figure*}
\centering
\includegraphics[width=\hsize]{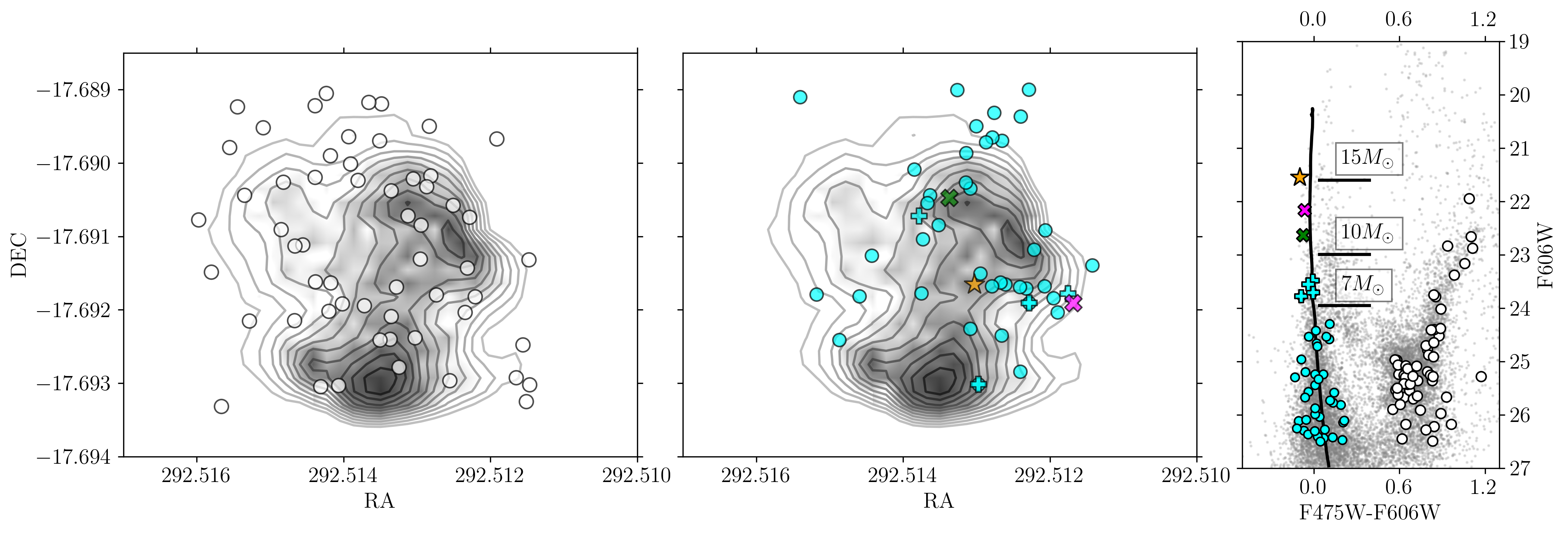}
\caption{Comparison of the SagDIG \ion{H}{ii} region H$\beta$ emission (black contours) with the stellar HST photometry of \citet{momany05}. The left panel shows the old RGB and red clump stars (white dots) in the \ion{H}{ii} region. The middle panel shows the distribution of young MS stars ($<7M_{\odot}$ and $7-10M_{\odot}$, cyan dots and cyan crosses, respectively) in the \ion{H}{ii} region, the orange star is the most luminous ($15 M_{\odot}$), and the magenta and green crosses are the second and third ($>10M_{\odot}$) most luminous stars. The right panel shows the CMD of SagDIG, with the same symbols and colours as the left and the middle panel. The vertical black curve is a 10 Myr isochrone showing the MS of young stars to get a proxy of the stellar masses of the young stars. The horizontal black lines show the corresponding range of masses in the main sequence.}
\label{fig12}
\end{figure*}

\citet{momany05} by using HST photometry to explore SagDIG's stellar populations reported a young main sequence, being the result of an ongoing star-formation episode, and a well-defined red giant branch, indicating that SagDIG also had an extended star-formation episode between 1-9 Gyrs ago. The suggestions from those features were confirmed by SFHs derived by \cite{held07}, and \cite{parto23}. The old stars are uniformly distributed across the SagDIG structure. On the other hand, young MS stars are located closer to the high \ion{H}{i} gas density clump previously studied by \cite{younglo97}.
\\
\\
To analyse the spatial distribution of stars located inside the \ion{H}{ii} region, we applied cuts on the SagDIG's CMD similar to those of \citet{momany05}, as is shown in the right panel of Figure \ref{fig12}: old RGB stars and red clump stars ($\sim 1-9$ Gyr) as $0.5<F475W-F606W<1.2$ (white dots), and young MS stars ($31-630$ Myr) as $F475W-F606W < 0.3$ and $F606W<26.5$ (cyan dots and crosses), where the most luminous MS star is shown with the orange star, and the second and the third ones are shown with magenta and green crosses, respectively.

The spatial distribution of the selected old RGB and red clump stars in the \ion{H}{ii} region (black contours) is shown in the left panel of Figure \ref{fig12}. Those stars are uniformly distributed. However, the young MS stars do not follow the same trend, as is seen in the middle panel. Those MS stars are, located closer to the edges of the Southern and the North-West H$\beta$ clumps in a filamentary-like configuration, and the most luminous star is located nearly the centre of the biconic-like distribution. 

Stars are born in high gas-density regions, where the ionizing flux of young massive stars expels the surrounding gas. This results in either denser gas regions with stars located in-situ, and/or crowded regions of stars surrounded by an ionized gas structure. Therefore the spatial distribution shown in the middle panel of Figure \ref{fig12} is not expected.

The spatial distribution of young MS stars could be produced by an observational bias: \textcolor{black}{the clumps of H$\beta$ emission may be potential regions of active star formation, so it is possible that those zones could contain significant dust that prevents us to detect young stars, i.e., high $C\mathrm{(H\gamma)}$ values (or high H$\beta$/H$\gamma$ ratio) compared with the remaining zones of the \ion{H}{ii} region.} However, the reddening constant $C(\mathrm{H\gamma})$ shows a flat behaviour across the \ion{H}{ii} region. Therefore, the spatial distribution of the SagDIG stars in the \ion{H}{ii} region must be due to a physical phenomenon. \\

As is shown in the right panel of Figure \ref{fig12}, a 10 Myr PARSEC isochrone \citep{bressan12} was used, adopting Z $= 0.004$ and $A_{V}=0.62$ \citep{momany05} to get a proxy of stellar masses of the young MS stars, suggesting that the most luminous star has a mass of $15M_{\odot}$ (orange star), and the second and third most luminous stars have masses in the range of $10-15M_{\odot}$. The middle panel of Figure \ref{fig12} shows that the $15M_{\odot}$ is located near the centre, whereas the two stars between 10 and $15M_{\odot}$ are located at the edges of the biconic-like shape.

\begin{figure*}
\centering
\includegraphics[width=\hsize]{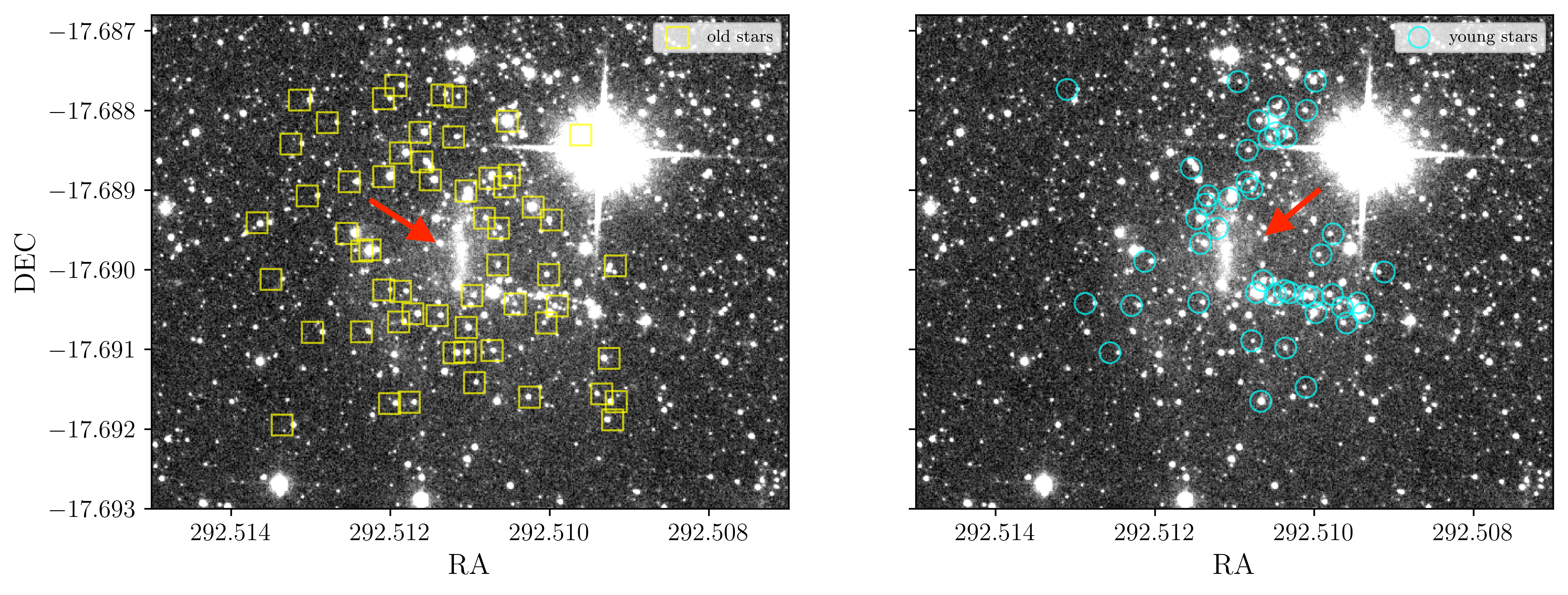}
\caption{HST image (F475W, F606W and F814W combined) acquired by \citet{momany05} observations, showing the SagDIG \ion{H}{ii} region. Left panel shows the location of the old stellar population with yellow squares. Right panel shows the young stellar population with cyan circles. In both panels, the vertical diffuse column density is indicated with the red arrow. The source in the upper right region is a foreground star.}
\label{fig12b}
\end{figure*}

In order to better understanding the SagDIG \ion{H}{ii} region, we visually inspected the HST image acquired from \citet{momany05}, shown in Figure \ref{fig12b}. The left panel shows the distribution of the old stellar population with yellow squares, whereas the right panel shows the distribution of the young stellar population with cyan circles. As is seen, the horizontal filamentary structure is clearly seen in the right panel, and the diagonal filamentary structure seems to be a mix of clumps of stars aligned. In addition, there is a large vertical emitting structure likely formed by young stars not resolved, which are not in the \citet{momany05} photometric catalogue.
\\
\\

OB-type MS stars are able to ionise their surrounding gas by radiation pressure in UV wavelengths, where the range of observed wind velocities in OB stars goes up to a few hundred km s$^{-1}$ (\citealt{lamers99}, \citealt{lamers17}, \citealt{osterbrock06}). However, there are stars with masses $<10M_{\odot}$, covering ages up to 630 Myr \citep{momany05} located at the edges of the lobes, suggesting that the two-lobe structure was created before the ionising stars in the centre were born. This feature is also in line with observational studies which suggest that stellar outflows can trigger sequential star formation, setting age gradients across \ion{H}{ii} regions (\citealt{matsuyanagi06}, \citealt{ogura07}, \citealt{rygl14}). Hence, a mix between SNe explosions and UV ionisation from massive stars is likely shaping the evolution of the  \ion{H}{ii} region. %done this one

Because the SagDIG \ion{H}{ii} region is expanding, and it is located at the edge of the densest \ion{H}{i} density clump (\citealt{younglo97}), the interaction between the \ion{H}{i} component and the gas ionised from massive stars and SNe explosions could somehow create the filamentary structures of young stars. \ion{H}{ii} regions show complex morphologies depending on ionising sources, non-uniform gas distributions, and young stellar objects are frequently detected in layers and filamentary gas structures, such as those found in the Orion Nebula, the Eagle Nebula, the Rosette Nebula and other galactic \ion{H}{ii} regions (\citealt{hill12}, \citealt{tremblin14}, \citealt{zavagno20}, \citealt{gaudel23}). The existence of filamentary structures has been known since the last century (\citealt{schneider79}, \citealt{ungerechts87}), but only \textit{Herschel} observations were able to link the existence of those objects with star formation (\citealt{andre10}, \citealt{molinari10}). However, although their properties are well identified, they are poorly constrained. High-resolution observations will help to get an understanding of their formation and evolution, putting new constraints in the underlying physics employed by theoretical simulations (e.g., \citealt{xu19}).\\
\\
The interaction between the \ion{H}{i} material and \ion{H}{ii} regions for the formation of filaments is still an open question (see discussion in \citealt{zavagno20}), but important clues were given by \cite{tremblin14}, which demonstrate that the ionisation feedback (compression of the material induced by the expansion of ionized gas) is key in assembly material in the photo-dissociation region (PDR) to form filamentary structures. Theoretically, the turbulent gas compressed by UV radiation from OB stars may trigger star formation (\citealt{menon20}). However, few observational works suggest that the net effect of the ionised radiation and stellar winds can trigger star formation (\citealt{billot10}, \citealt{chauhan11}, \citealt{roccatagliata13}).

\subsection{The SagDIG \ion{H}{ii} region in the mass-metallicity plane}
Analyzing the physical structure of the \ion{H}{ii} region with both the VIMOS-IFU and the EFOSC2-long slit data gave us important insights into the physical mechanisms that are shaping the evolution of the gaseous nebula. But they can also be constrained by the gas-phase chemical content. Therefore, we proceed to place the SagDIG \ion{H}{ii} region in the mass-metallicity plane. It is important to note, that we will use the total stellar mass of SagDIG, and assume that the metallicity of the only known \ion{H}{ii} region is representative of the whole galaxy. Therefore, the inferred properties should be interpreted with caution, considering the potential uncertainties and the limited sample size.

\begin{figure}
\centering
\includegraphics[width=\hsize]{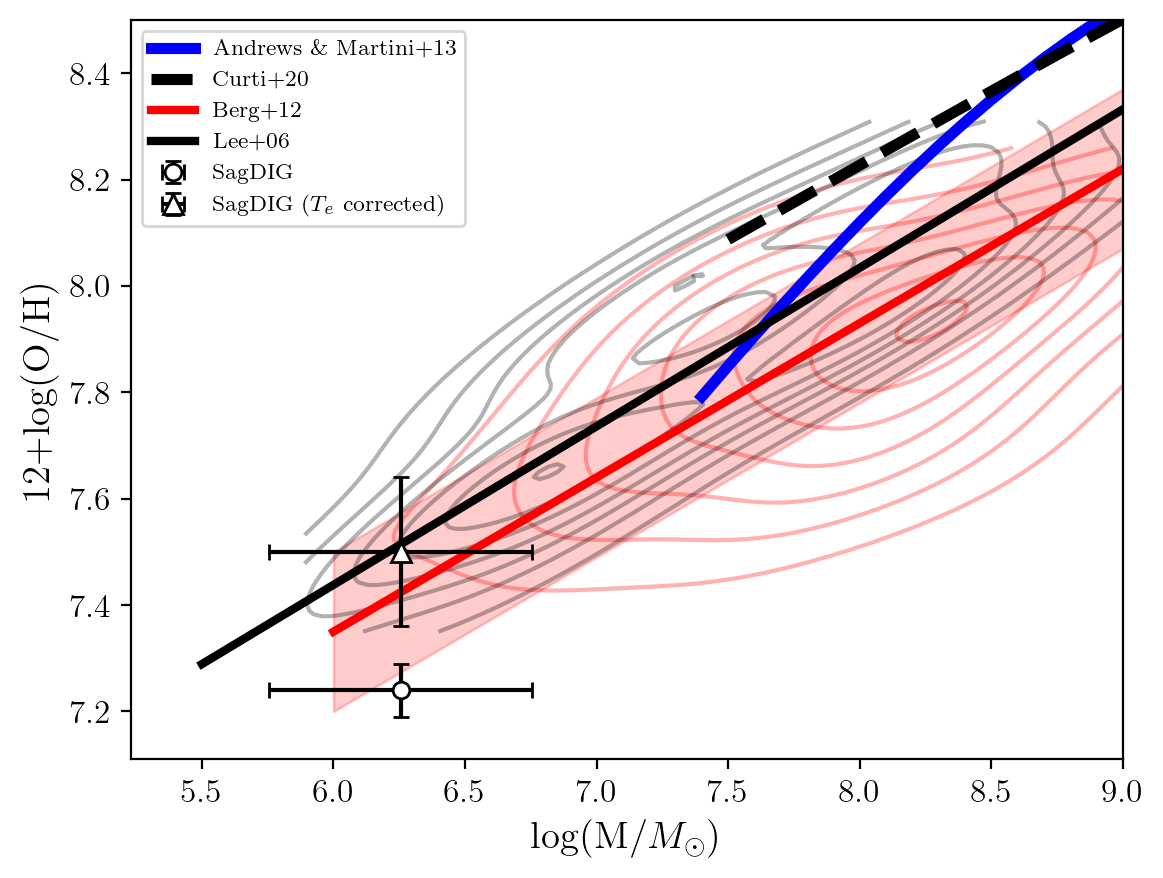}
\caption{SagDIG in the mass-metallicity plane shown white triangle and white dots, for the $T_{e}$ corrected and uncorrected oxygen abundance, respectively. Black dotted and blue solid lines are the local universe $T_{e}-$based MZR of \citet{curti20} and \citet{andrewsmartini13}. Black and red solid lines are the low-mass end of the MZR of \citet{lee06} and \citet{berg12}. Contours are the respective galaxy sample of dwarf irregular galaxies used to fit the low-mass end of the MZR with the corresponding colors, and the red dashed area is the respective scatter of $\sigma=0.15$ dex, of the \citet{berg12} low-mass end of the MZR.}
\label{fig13}
\end{figure}

We used $1.8\pm 0.5\times 10^{6}M_{\odot}$ as the stellar mass of SagDIG from \citet{kirby18}. This value is based on the stellar mass-to-light ratios of \citet{woo08}. Because of the biases discussed in Section \ref{Te-biases}, we consider both the $T_{e}$ corrected and the uncorrected gas-phase metallicities of the SagDIG \ion{H}{ii} region (from the VIMOS-IFU data), since most of the works that employ this method do not take into account this bias. Figure \ref{fig13} shows SagDIG in the mass-metallicity plane with the white triangle and white dot as the $T_{e}$ corrected and uncorrected metallicities, respectively. We compared those values with the $T_{e}-$based local universe MZR of \citet{andrewsmartini13} and \citet{curti20}. In addition, we also compare our results with the low-mass end of the MZR of \citet{lee06} and \citet{berg12}. Colour contours represent the galaxy sample of dwarf irregular galaxies used to fit the low-mass end of the MZR in both works.

The intrinsic scatter of the MZR increase as the metallicity and mass decrease (\citealt{zahid12}), because low-mass galaxies are more likely to be disturbed by pristine gas inflow and outflow mechanisms, in which these gas flows in those systems can change the efficiency of the gas to form stars (\citealt{peeples08}, \citealt{dave11}, \citealt{lilly13}). Therefore, the non-equilibrium regime will put a galaxy upward/downward in the mass-metallicity plane with respect to the main locus end of the MZR. In particular, outflow processes from SNe explosions in low-mass galaxies have a significant impact because they act on shallower potential wells. Hence, it is inherent to conclude that, in general, the evolution of a low-mass metal-poor galaxy is truncated when outflow mechanisms are acting, expelling metals with a high mass-loading factor \citep{chisholm18}. 

Zoom-in hydrodynamical simulations demonstrate that small galaxies ($M_{dyn}<10^{8}M_{\odot}$) grow in stellar mass up to $z\sim4$ as a consequence of cosmic reionisation, where the strong stellar feedback driven by galactic winds increase the ISM temperature, preventing the gas infall onto those galaxies (\citealt{agertz20}, \citealt{rey24}). This results in a slow chemical evolution in dwarf galaxies, which is in line with observational evidence of suppression of star formation at early epochs in dwarf galaxies (\citealt{held07}, \citealt{cole07}, \citealt{tolstoy09} ). Hence, the shallower slope of the low mass end of the MZR starts to be observed from the high-redshift universe ($z>6$, \citealt{curti23}) to the local universe (\citealt{lee06}, \citealt{berg12}), likely shaped by galaxies being affected by outflows triggered by momentum-driven SNe winds according to the theoretical models by \citet{murray05}, and  \citet{dave12}.

As viewed in Figure \ref{fig13}, the location of SagDIG in the mass-metallicity plane falls inside the $1\sigma$ scatter (0.15 dex) of the low-mass end of the MZR, suggesting that the SagDIG \ion{H}{ii} region is likely being dominated by low-scale outflows processes, being in line with the features observed in the previous subsections.
\\
\\
The observed stratification of the ionised gas, revealed by the emission line maps and flux-density profiles, the filamentary configuration of the young stellar populations, together with their low-metallicity content which places SagDIG in the low-mass end of the MZR, suggest that the scenario that best explains the observed features of the SagDIG \ion{H}{ii} region, is that the gaseous structure is being dominated by stellar feedback processes, such as ionisation coming massive stars, SNe explosions, and stellar winds. This would mean, that extragalactic \ion{H}{ii} regions in the local universe are being governed by the similar underlying physics as \ion{H}{ii} region in the Milky Way.
 
This proposed scenario should be tested with (i) kinematic analysis of the ionized gas, (ii) exploring other atomic species (such as $\mathrm{N^{+}}$, and $\mathrm{S^{+}}$), (iii) Infrared observations to study the dust content and the filamentary structures, and (iv) UV observations to get information of stellar winds coming from massive OB stars. Instruments, e.g., MUSE/VLT, NIRSpec/JWST, and MIRI/JWST will help to disentangle the nature of the evolution of the SagDIG \ion{H}{ii} region with better spatial resolution and spectral coverage. 

\section{Summary and conclusions}
This work presents, for the first time, a detailed analysis of an \ion{H}{ii} region beyond the Milky Way.

We used archival optical, intermediate-resolution VIMOS/VLT-IFU and EFOSC2/NTT-long slit data to explore the physical structure of the only known \ion{H}{ii} region in SagDIG, to probe the physical mechanisms which are shaping the chemical evolution of this nebula. We detected the auroral [\ion{O}{iii}]$\lambda4363$ emission line and estimate oxygen abundances with the direct method, integrating the spectral fibres covering by the \ion{H}{ii} region. The IFU data allow us to generate emission line maps and flux-density profiles of H$\beta$ and [\ion{O}{iii}]$\lambda5007$ to study their structures. Furthermore, the EFOSC2 long-slit data allow us to explore emission line flux variations in the East-West direction of the \ion{H}{ii} region with more emission lines than the IFU data. 

The analysis done in this work shows that the \ion{H}{ii} region of the metal-poor dwarf galaxy SagDIG shows a similar evolution as galactic \ion{H}{ii} regions. Our main conclusions are summarised as follows: 
\begin{itemize}
    \item The SagDIG \ion{H}{ii} region shows two prominent H$\beta$ clumps. The flux-density maps reveal that those clumps are similar between them in terms of size and fluxes, in H$\beta$ and [\ion{O}{iii}]$\lambda5007$ emission lines. Those also seem to be aligned across the same axis.
    \item The $\mathrm{O^{++}}$ distribution is more concentrated towards the centre, in comparison with low ionisation species such as $\mathrm{O^{+}}$ and $\mathrm{H^{+}}$, suggesting that the stratified composition of this nebula is due to ongoing expansion. 
    \item The young stellar population is located closer to the edges of the two clumps, distributed in a filamentary-like configuration, whereas the old stellar population is uniformly distributed across the \ion{H}{ii} region. Those features suggest that stellar feedback mechanisms, such as UV radiation from massive stars, stellar winds, and SNe explosions are likely the main physical phenomena that are shaping the evolution of this nebula.
    \item The \textcolor{black}{direct method $T_{e}-$based} oxygen abundance of the SagDIG \ion{H}{ii} region is \textcolor{black}{$\mathrm{12+log(O/H)}=7.23 \pm 0.04$, or $\mathrm{12+log(O/H)}=7.50 \pm 0.08$} applying the \citet{cameron23} temperature correction. 
    \item The SagDIG \ion{H}{ii} region shows spatial variations in terms of decreasing $T_{e}$ and increasing $T_{e}-$based oxygen abundances of \textcolor{black}{$\sim$1500 K and $\sim$0.25} dex towards the outskirts, respectively, in line with the stratified composition.
    \item The position of SagDIG in the mass-metallicity plane is consistent with the low-mass end of the MZR, suggesting that the \ion{H}{ii} region is dominated by stellar feedback mechanisms. 
\end{itemize}

This proposed scenario can be tested with high-performance instruments, by acquire IFU data with better spectral coverage and spatial resolution, together with infrared spectro-photometric analysis.

%--------------------------------------------------------------------
\begin{acknowledgements} 
We thank the anonymous referee for their useful comments which improved the quality of the paper. A special thanks goes to Eric Andersson, Joseph Anderson, Bastian Ayala, Ana Jimenez, Timo Kravtsov, Thomas Moore, and Felipe Vivanco Cádiz for the opportunity to discuss our results. \textcolor{black}{L.M. acknowledges support from ANID-FONDECYT Regular Project 1251809.}
\end{acknowledgements}

\begin{appendix}

\section{Combination of EFOSC2 long-slit and VIMOS-IFU mock-slit spectrum for $T_{e}-$based oxygen abundance estimations}
\label{ap1:metallicities}

\begin{table*}[h!]
\label{tableA}
\caption{Electron temperature and ionic oxygen abundances estimate for SagDIG using Pyneb. The second columns shows the $T_{e}$ estimates, the third and fourth columns shows the ionic oxygen abundances $\mathrm{O^{+}/H}$ and $\mathrm{O^{++}/H}$, respectively. The fifth and sixth columns shows the total oxygen abundances with the $T_{e}-$sensitive method, and the $R_{23}$ empirical calibrator, respectively.}
\centering
\begin{tabular}{cccccc}
\hline \hline
                                    & $T_{e}$ & $12+\mathrm{log(O^{+}/H)}$ & $12+\mathrm{log(O^{++}/H)}$ & $12+\mathrm{log(O/H)}$ & $12+\mathrm{log(O/H)}_{R_{23}}$ \\ \hline
VIMOS mock slit - Pyneb             & $22165\pm 1122$K&   $6.98\pm 0.02$& $ 6.83\pm 0.06$ & $7.23\pm 0.04$ & $7.31 \pm 0.08$\\
EFOSC2 S02 - Pyneb & --- &                               $6.98\pm 0.02$&  $6.82 \pm 0.02$ & $7.22 \pm 0.03$& $7.30 \pm 0.10$ \\

\hline

\hline

\end{tabular}
\end{table*}

\textcolor{black}{The total oxygen abundance is defined as $12+\mathrm{log(O/H)}$, where $\mathrm{O/H = (O^{+}/H^{+} + O^{++}/H^{+})}$. The VIMOS-IFU data allow us to measure the $\mathrm{O^{++}/H}$ ionic abundance estimate. However, the VIMOS-IFU spectral range does not cover the wavelength region where $[\ion{O}{ii}]\lambda 3727$ is found. For this reason, we used the EFOSC2 long-slit spectrum from S02 to measure the $[\ion{O}{ii}]\lambda 3727$ for a posterior $\mathrm{O^{+}/H}$ ionic abundance estimate.} 

\begin{figure}
\centering
\includegraphics[width=\hsize]{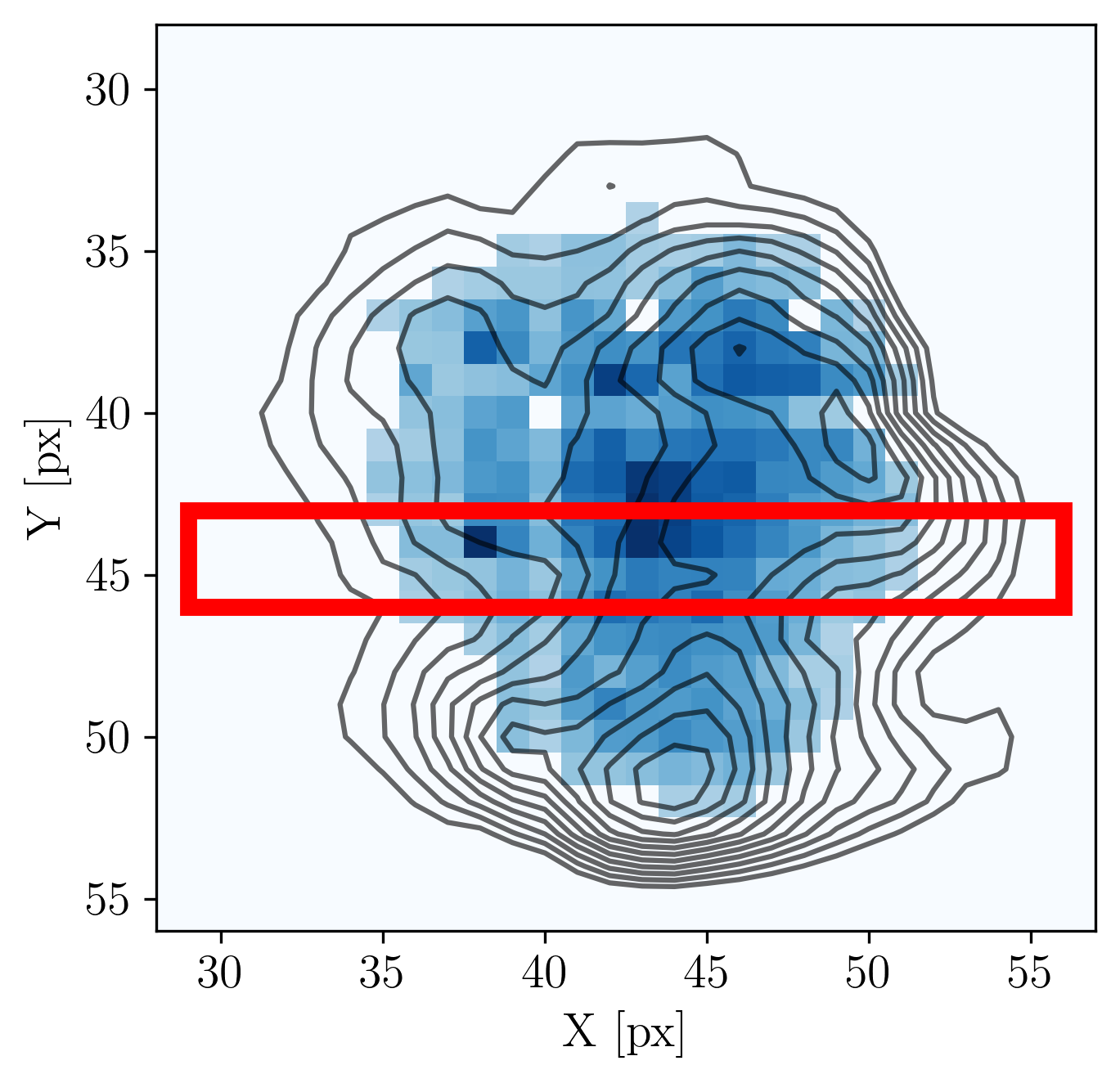}
\caption{{Simulated slit in the IFU data following the EFOSC2 long-slit observations of S02, shown with the red rectangle. The blueish pixels are those selected to generate the integrated SagDIG spectrum shown in Figure \ref{fig2}, with their respective jump value shown in the color bar. The black contours are represent the H$\beta$ emission line map.}}
\label{fig8aaaaa}
\end{figure}

\textcolor{black}{The combination of the VIMOS-IFU integrated spectrum and the EFOSC2 long-slit spectrum should introduce instrumental biases since those datasets come from different instruments covering different areas of \ion{H}{ii} region. We face this problem by simulating a slit in the VIMOS data cube following the S02 observing setup, shown with the red rectangle in Figure \ref{fig8aaaaa}. We have also plotted the spectral fibres selected to generate the integrated spectrum with blueish colours (same as the right panel of Figure \ref{fig2}), and the H$\beta$ emission map of the SagDIG \ion{H}{ii} region was superimposed for reference.}

\textcolor{black}{We found that the reddening corrected emission line measurements of S02 (Table 1 of S02) present a emission line ratio H$\gamma$/H$\beta = 0.41 \pm 0.11$. H$\gamma$/H$\beta$ should be 0.468 under Case B recombination, $T_{e}=10000$ K, and $N_{e}=100$ $cm^{-3}$. This may be produced because S02 perform dust correction with the theoretical ratio H$\alpha$/H$\beta = 2.85$ in the EFOSC2 grism GR$\#$9 spectrum, producing a flux difference for the grism GR$\#$7 spectrum, where the emission lines selected to estimate $R_{23}-$based metallicities are taken. For this reason, we proceed to perform dust-corrected flux measurements using the  H$\gamma$/H$\beta = 0.468$ theoretical ratio in the EFOSC2 GR$\#$7 spectrum.} \\
\\
\textcolor{black}{To test reliability in the combination between those two spectra for estimating $T_{e}-$based metallicities, we first compare total oxygen abundances between the mock-slit spectrum and the EFOSC2 spectrum via strong line method $R_{23}$ \citep{kk04}, being the same as used in S02. The $R_{23}$ index is defined as $R_{23} = [I(3727)+I(4959)+I(5007)]/I(\mathrm{H\beta})$.} \\
\\
\textcolor{black}{We combine the $I(3727)$ measurement from the EFOSC2 long-slit and $I(4959)+ I(5007)$ from the VIMOS mock-slit, under the same theoretical ratio to apply dust-correction (H$\gamma$/H$\beta = 0.468$). The $R_{23}$ index with this combination is derived as follows:}

\begin{equation}
    \textcolor{black}{R_{23} = \left( \frac{I(3727)}{H\beta}\right)_{\mathrm{EFOSC2}} + \left( \frac{I(4959)+I(5007)}{H\beta}\right)_{\mathrm{VIMOS\ mock-slit}}}
\end{equation}
\textcolor{black}{On the other hand, $R_{23}$ index using all the required emission lines from EFOSC2 is estimated with the following expresion:}
\begin{equation}
    \textcolor{black}{R_{23} = \left( \frac{I(3737)+I(4959)+I(5007)}{H\beta} \right)_{\mathrm{EFOSC2}}}
\end{equation}
\\
\textcolor{black}{So if the derived $R_{23}-$based metallicities are in agreement between those two ways to estimate $R_{23}$, we are able to combine the flux measurements to estimate $T_{e}-$based oxygen abundances, since the combination relies on the flux measurements under the same theoretical value in the dust reddening corrections.}\\ 
\\
\textcolor{black}{We derived a $R_{23}-$based oxygen abundance as $12+\log(\mathrm{O/H})= 7.30 \pm 0.10$ using all the required emission line from the EFOSC2 spectrum. Then, using $I(3727)$ form the EFOSC2 and $I(4959)+I(5007)$ from the VIMOS mock-slit, we derived a $R_{23}-$based oxygen abundance as $12+\log(\mathrm{O/H})= 7.31 \pm 0.08$. Those values are in excellent agreement.}\\
\\
\textcolor{black}{Hence, we proceed to estimate $T_{e}$ for a subsequent derivation of the oxygen abundance using the direct method. We estimate $T_{e}[\ion{O}{iii}] = 22165\pm1122$ K, and $T_{e}[\ion{O}{ii}] = 18515 \pm 987$. The total oxygen abundances for the VIMOS-mock slit is $12+\mathrm{log(O/H)} = 7.23 \pm 0.04$, and the total oxygen abundance using the EFOSC2 measurements results in $12+\mathrm{log(O/H)} = 7.22 \pm 0.03$. Those estimates are in agreement. The values estimated with this procedure are shown in Table \ref{tableA}. }\\
\\
\textcolor{black}{Uncertainties were computed by running 1000 Monte Carlo simulations. We represent each emission line as a random value of a Gaussian distribution centred on the line flux measurement and a standard deviation as the error of the flux measurement.}

\end{appendix}
\end{document}